\documentclass[10pt,letterpaper]{article}
\usepackage[top=0.85in,left=2.75in,footskip=0.75in]{geometry}

\usepackage{amsmath,amssymb}

\usepackage{changepage}

\usepackage{textcomp,marvosym}

\usepackage{cite}

\usepackage{nameref,hyperref}

\usepackage{algorithm}
\usepackage{algpseudocode}
\usepackage{comment}

\usepackage{multirow}

\usepackage[right]{lineno}

\usepackage[nopatch=eqnum]{microtype}
\DisableLigatures[f]{encoding = *, family = * }

\usepackage[table]{xcolor}

\usepackage{array}

\newcolumntype{+}{!{\vrule width 2pt}}

\newlength\savedwidth

\newcommand\thickhline{\noalign{\global\savedwidth\arrayrulewidth\global\arrayrulewidth 2pt}%
\hline
\noalign{\global\arrayrulewidth\savedwidth}}

\raggedright
\usepackage[aboveskip=1pt,labelfont=bf,labelsep=period,justification=raggedright,singlelinecheck=off]{caption}

\makeatletter
\renewcommand{\@biblabel}[1]{\quad#1.}
\makeatother

\usepackage{lastpage,fancyhdr,graphicx}
\usepackage{epstopdf}
\fancyheadoffset[L]{2.25in}
\fancyfootoffset[L]{2.25in}
\begin{document}
\vspace*{0.2in}


\begin{flushleft}
{\Large
\textbf\newline{UpStory: the Uppsala Storytelling dataset} 
}
\newline
\\
Marc Fraile\textsuperscript{1*},
Natalia Calvo-Barajas\textsuperscript{1},
Anastasia Sophia Apeiron\textsuperscript{2},
Giovanna Varni\textsuperscript{3},
Joakim Lindblad\textsuperscript{1},
Nataša Sladoje\textsuperscript{1},
Ginevra Castellano\textsuperscript{1}
\\
\bigskip
\textbf{1} Department of Information Technology, Uppsala University, Uppsala, Sweden
\\
\textbf{2} Department of Information and Computing Sciences, Utrecht University, Utrecht, Netherlands
\\
\textbf{3} Department of Information Engineering and Computer Science, University of Trento, Trento, Italy
\\
\bigskip

* marc.fraile.fabrega@it.uu.se

\end{flushleft}

\section*{Abstract}

Friendship and rapport play an important role in the formation of constructive social interactions, and have been widely studied in educational settings due to their impact on student outcomes. Given the growing interest in automating the analysis of such phenomena through Machine Learning (ML), access to annotated interaction datasets is highly valuable. However, no dataset on dyadic child-child interactions explicitly capturing rapport currently exists. Moreover, despite advances in the automatic analysis of human behaviour, no previous work has addressed the prediction of rapport in child-child dyadic interactions in educational settings. We present UpStory --- the Uppsala Storytelling dataset: a novel dataset of naturalistic dyadic interactions between primary school aged children, with an experimental manipulation of rapport. Pairs of children aged 8-10 participate in a task-oriented activity: designing a story together, while being allowed free movement within the play area. We promote balanced collection of different levels of rapport by using a within-subjects design: self-reported friendships are used to pair each child twice, either minimizing or maximizing pair separation in the friendship network. The dataset contains data for 35 pairs, totalling 3h 40m of audio and video recordings. It includes two video sources covering the play area, as well as separate voice recordings for each child. An anonymized version of the dataset is made publicly available, containing per-frame head pose, body pose, and face features; as well as per-pair information, including the level of rapport. Finally, we provide ML baselines for the prediction of rapport.


\section*{Introduction}

Rapport, the establishment of a close relationship based on mutual understanding, has long been considered to be an important facet of social interactions~\cite{tickle1990nature}, and particularly in dyadic (one-on-one) interactions~\cite{travelbee1963we}. In the educational context, teacher-student rapport and student-student rapport have been shown to have a positive impact on the learning experience~\cite{frisby2010instructor}. Friendship between classmates is naturally a high-rapport relationship and has been shown to improve measurable outcomes such as task performance, effective problem-solving, and learning outcomes~\cite{azmita1993friendship, newcomb1995children, wentzel2018do}. 

While the ML-based analysis of affect and related interpersonal constructs has been well studied (see e.g.~\cite{wu2020automatic, alsofyani2021attachment, nojavanasghari2016emoreact}), no previous work has investigated the prediction of rapport in child-child dyadic interactions in educational settings. Moreover, no dataset capturing this interaction modality is available. The primary goal of this study is to address these gaps by introducing a publicly available child-child interaction dataset annotated for rapport, and providing ML baselines for its automatic prediction.

In order to learn useful models of social interaction through ML techniques, practitioners require access to high-quality multimodal datasets. Due to the sensitive nature of the recorded data (including the likeness and voice of the participants, as well as the information they disclose through speech), the number of such datasets is relatively small, and access is constrained. Some well-known datasets are available for adult-adult interactions~\cite{bilakhia2015the} and child-child interactions~\cite{singh2018p2pstory,lemaignan2018pinsoro}, but in these, available annotations for social constructs are obtained from external raters. This has implications on the reliability of the annotations, and an objective measure of rapport would be preferable when studying this phenomenon. We address this issue by relying on the \textit{friendship network}: a graph representing friendships in the classroom through self-reported friendship nominations. 
Friendship networks have long been used to study the social relationships in educational institutions~\cite{hallinan1979structural}, and have been analyzed using graph theory to evaluate the social fabric of the student population~\cite{hansell1985adolescent}. In this study, we propose a partition scheme based on friendship network analysis to obtain high-rapport and low-rapport pairings of the student population. We validate the partition scheme using graph analysis and questionnaire responses.

The data collection effort is centered on an educational collaborative game designed to elicit pair interactions. Pairs of children determined with the proposed partition scheme are invited to play a card-based storytelling game together. They plan a story in a private space, while being recorded by two overhead cameras and head-mounted microphones. Each pair plays several rounds of the game; the obtained recordings form a private multimodal dataset. The UpStory dataset is subsequently obtained by extracting anonymized face and pose features from the video recordings. Finally, we set ML baselines for the prediction of rapport, either using features from a single child, or considering the joint data from both children in a game round.

In summary, the main contributions of this paper are: 
\begin{enumerate}
    \item the proposal of a novel experimental manipulation based on friendship network analysis to obtain high-rapport and low-rapport pairings of the student population in educational settings;
    \item the collection of a new multimodal dataset on child-child interaction, with an experimental manipulation for the level of rapport, including recordings of children interacting in a task-oriented activity, captured from two audio sources (head-mounted microphones) and two video sources (overhead cameras);
    \item the publishing of UpStory, an open access feature dataset; and
    \item the establishment of ML baselines for the prediction of rapport in the feature dataset.
\end{enumerate}

The UpStory dataset is publicly available at \url{https://zenodo.org/doi/10.5281/zenodo.12635620}

The source code for this project is available at \url{https://github.com/MarcFraile/dyadic-storytelling}

\section*{Related Work}

\subsection*{Children's rapport}
\label{sec:rw:rapport}

Rapport requires two individuals willing to make and establish a meaningful connection. As rapport occurs when people’s communication styles synchronize, it facilitates communication during task-oriented activities, fostering a higher degree of cooperation, and leading to a more profound level of engagement~\cite{newcomb1995children, azmita1993friendship, wentzel2018do}. Children’s relationships with others evolve as they progress through various developmental stages impacting their cognitive development. Engaging in discussions and collaborative tasks with friends and peers prompts children to explore and critically evaluate varying perspectives, contributing to their cognitive growth~\cite{hartup1996thecompany, bukowski1998company}. For these positive effects to materialize, establishing relationships characterized by quality is crucial. In this context, rapport-building becomes essential in assessing children’s positive connections with others, involving mutual understanding and caring~\cite{tickle1990nature}.  

Behaviors such as cooperation, social contact, positive affect, and verbal communication are significantly more prevalent among individuals who share rapport~\cite{tickle1990nature}. Numerous studies have aimed to develop tools for predicting and identifying peer relationships, employing diverse methodologies such as direct observation~\cite{berndt1988friends}, manual video annotation~\cite{ladd1990preschoolers}, sensor-based assessment~\cite{kanda2006anapproach}, and automatic prediction techniques~\cite{messinger2022computational}. These approaches are based on analyzing group behaviors, where social relationships are evaluated through measurements of member distances, facial expressions, conversational patterns, synchrony, and time spent together~\cite{messinger2022computational, lin2023friends, rabinowitch2017synchronized, maman2020game}.

While those studies have utilized diverse approaches to analyze group behaviors and evaluate social relationships, there is a need to understand the broader context of peer interactions. As such, research in Child Development has emphasized the exploration of social networks to understand children's status within a social group~\cite{kasari2011social}. More specifically, friendship networks provide insights into the dynamics of peer relationships and the formation of social bonds and rapport. Approaches such as self-reported friendship nominations contribute to the identification of social networks by identifying key players and patterns of connections~\cite{george1996friendship, strauss2003social}. This research aims to provide a dataset for the automatic prediction of rapport in child-child dyadic interactions. Therefore, we relied on \textit{social networks} to identify dyads with high and low rapport and, consequently, to comprehend the dynamics in these dyads. 

\subsection*{Datasets on Dyadic Interactions}
\label{sec:int:datasets}

Dyadic social interaction datasets are crucial for understanding how individuals engage in various activities and gaining insights into interaction dynamics. For instance, the IEMOCAP dataset~\cite{busso2008iemocap} consists of audiovisual recordings of English-speaking actors, either (1) acting out a scene or (2) improvising on a pre-defined topic. The dataset aims to elicit emotional expressions and provides annotated data for different categories of emotions such as happiness, anger, and sadness, and labels for valence and dominance dimensions. Another dataset is ALICO~\cite{malisz2016alico}, which consists of audiovisual recordings of German-speaking adults taking either the role of a storyteller (sharing a vacation story) or an active listener. The ALICO dataset provides hand annotations for head movements and the function of the listener's responses, among others, characterizing mechanisms in interpersonal communication. 

While these datasets provide great value to the study of social interactions between dyads, they suffer from low agreement between raters when annotating the data, which could affect the reliability and quality of it~\cite{alhazmi2021effects}. Cohen's Kappa is a statistical criterion used to measure the reliability between raters, for the ALICO dataset, this value is around $\kappa = 0.30$, which is considered low~\cite{mchugh2012interrater}. Moreover, the majority of these existing datasets focus on examining dyads with adults as the target population with a limited focus on capturing children’s social interactions~\cite{busso2008iemocap, malisz2016alico, bilakhia2015the}. The lack of available datasets to study children's interactions makes it difficult to generalize findings across different age groups and understand the developmental aspects of social interactions.

Datasets designed to capture children's social interactions often emphasize the adult-child context. These datasets have primarily concentrated on evaluating children's reactive emotions to objects \cite{nojavanasghari2016emoreact}, exploring play in the context of autism diagnoses \cite{rehg2013decoding}, or annotating and estimating synchrony between a child and a therapist \cite{li2021improving}. Recent efforts have introduced a comprehensive multimodal dataset capturing the dynamics of parent-child interaction within a storytelling context \cite{chen2023dyadic}. However, the tasks selected to capture social interactions in these datasets do not encompass the natural interactions that arise when children engage with their peers. 

Only a few datasets focus on child-child interactions. The P2PSTORY dataset~\cite{singh2018p2pstory} focuses on child-child interactions during a collaborative storytelling task, with one child as the listener and the other as the storyteller. The dataset includes video and audio data, along with annotations of non-verbal behaviors such as gaze, smiles, forward leans, nods, and verbal cues like short utterances and backchanneling. Multiple coders manually annotated children's behaviors following a defined coding scheme. While adhering to standard annotation practices, coder agreement varied across behaviors, with more observable actions achieving higher agreement levels than subjective evaluations, highlighting the complexity of annotating behaviors with multiple label options.

The PInSoRo dataset~\cite{lemaignan2018pinsoro} is also dedicated to understanding the social dynamics among children in free-play interactions. This multimodal dataset provides video and audio recordings, employing automated computational methods for feature extraction, such as prosody, voice quality, facial landmarks, Action Units (AUs), gaze, and skeleton data. It also includes manual annotations of social interaction, focusing on task and social engagement, and social attitudes in child-child dyads and child-robot dyads. Similar to the P2PSTORY Dataset, PInSoRo's manual annotations involved multiple experts in human behavior; however, the challenge of coding social interaction is once again echoed, as it resulted in a low agreement between coders. 

While both datasets contribute to understanding the dynamics in children's interactions with others, they do not focus on the differences in interactions with friends versus acquaintances. This distinction is crucial, as interaction patterns can significantly vary based on the relationship between the individuals. Therefore, there is a need for datasets that explicitly categorize interactions based on the nature of the relationship, enabling a more nuanced analysis of children's social dynamics.

\subsection*{Automatic prediction of rapport}
\label{sec:int:CM}

Computational models in Affective Computing are emerging to better understand the complex dynamics of social interactions, such as analyzing and predicting human behavior, affect, and social constructs~\cite{wang2022systematic}. Therefore, different verbal and non-verbal behaviors could provide information on rapport in an interaction. For instance, the estimation of children's closeness to others using computational approaches has primarily been explored through location-based techniques, specifically by computing the distance separating two children to predict their friendship status. 
Kanda and Ishiguro implemented a friendship estimation algorithm by using wearable sensors' data to calculate potential friendships between children based on their interaction times, with specific attention to simultaneous interactions~\cite{kanda2006anapproach}. In another study, depth sensors' data was employed to design a tracking algorithm aimed at estimating the child's position. Additionally, RGB cameras were utilized for face identification. These data sources were subsequently integrated to provide an estimation of children's relationships with their peers within the classroom~\cite{komatsubara2018estimating}. 

In addition, synchrony and mimicry have been studied due to their impact on children's attitudes toward each other, enhancing social bonds with friends~\cite{denham2013iknow, rabinowitch2017synchronized}. For instance, in a peer-play context, a study involving children aged 4 to 6 years revealed that when interacting synchronously with their peers, children exhibited higher frequencies of helping behaviors, mutual smiles, and eye contact compared to asynchronous interactions~\cite{tuncgenc2018interpersonal}. Notably, mutual smiling and sustained eye contact were prominent features of interactions specifically among friends~\cite{goldman2007friendships}, suggesting that non-verbal behaviors play a crucial role when assessing children's attitudes and relationships in different social interactions. 

Other studies employ benchmark tools such as OpenFace~\cite{baltruvsaitis2016openface} for the extraction of facial features, OpenPose~\cite{cao2019openpose} for the extraction of body features, and OpenSMILE~\cite{eyben2010opensmile} for the extraction of voice features. Wu and colleagues conducted a user study in which university students participated in video call consultations with simulated patients~\cite{wu2020automatic}. Motor mimicry episodes were automatically detected using features extracted with OpenFace. Statistical analysis revealed a correlation between mimicry and communication skills in dyadic interactions. Alsofyani and Vinciarelli designed a multimodal approach for attachment recognition in children aged 5 to 9~\cite{alsofyani2021attachment}. Participants were asked to participate in a storytelling-based psychiatric test typically scored by an expert rater. In this case, an automated pipeline using OpenFace and OpenSMILE was used to predict the expert annotations, using a combination of logistic regression models for unimodal and multimodal data. Similar methodologies have been applied to analyze behaviors such as gaze duration and direction, and engagement, utilizing video data and exploring interactions between infants and children with adults~\cite{erel2021icatcher, fraile2021automatic}. However, the existing literature remains scarce in its examination of the dynamics of social interactions when assessing children's relationships with others. Consequently, tools that automatically predict the level of rapport based on children's behaviors are still limited. This limitation arises from the lack of appropriate datasets that distinguish between different types of rapport in children's dyads and the relevant features that enable accurate rapport prediction 

To address these gaps, we present the \textbf{UpStory} multimodal dataset, which focuses on capturing the dynamics of social interactions in children's dyads. We designed a study based on friendship network analysis to obtain low-rapport and high-rapport dyads. This method allowed us to study dynamic interactions between children and their peers in an educational context. Contrary to (subjective) manual annotation of the existing datasets, we used a friendship network methodology to obtain more objective annotations. In addition, we extracted behavioral features using benchmark tools such as OpenFace and OpenPose, and provided ML baselines for predicting the level of rapport in the dataset. By implementing the friendship network methodology, we ensured that the behaviors captured within the dataset were representative of social dynamics inherent to children's interactions, where rapport levels can vary significantly depending on their social status (i.e., the position that one holds in a group~\cite{coie1990peer}). As a result, the UpStory dataset facilitates a more robust analysis of children's social interactions and the automatic prediction of rapport.

\section*{Materials and Methods}

In order to collect the UpStory dataset, we designed a study with the explicit goal to collect samples of high-rapport and low-rapport pairs of children participating in a collaborative storytelling activity. This section details the design of the study, including the participant population, the pair-making strategy, and the primary data collection effort.

\subsection*{Participants}


The study was performed at a local primary school during free play hours. Prior to execution, the research plan was reviewed and approved by a local ethics committee\footnote{Etikprövningsmyndigheten, diarienummer 2022-04863-01.}. We collaborated with the school's teaching team to inform the students and their families about the study, and to collect consent forms signed by the children's legal guardians. Only children who presented a completed consent form and showed interest in participating were retained for the study. The children were informed that they could refuse to participate at any point, without consequences.

The recruited children were students in Year 2 (ages 8-9) and Year 3 (ages 9-10) who could speak English fluently. The collaborating school had a designated English-language class per year; students in the Swedish-language classes who were fluent in English were also allowed to participate. A total of 39 children signed up for the study and were assigned a two-digit random ID. One student was subsequently excluded due to lack of availability, resulting in a population of 38 participants: 28 students in Year 2 (14 boys and 14 girls), and 10 students in Year 3 (4 boys and 6 girls). 

\subsection*{Experimental Conditions}


\subsubsection*{Pair Making with a Social Distance Heuristic}
\label{sec:pair-making}

Since manual annotation of social constructs is 
prone to produce disagreement between coders~\cite{singh2018p2pstory,lemaignan2018pinsoro,fraile2022end}, it is preferable to capture high-rapport and low-rapport pairs through an explicit experimental manipulation. 
Given a large cohort size, a between-subjects design could be achieved by choosing pairs of participants that are either known to be close friends (high-rapport condition), or known to have separate social circles (low-rapport condition), at the cost of introducing selection bias. 
However, this is not a valid approach if we have access to a smaller cohort (such as classmates in a primary school), or if we wish to avoid the implicit bias.
To address this, we propose a pair-assigning strategy that is suitable for small cohorts, using a within-subjects design which ensures that each child participates once in each condition. Our method is based on relaxing the criteria for pair selection: we seek to optimize a \textit{social distance heuristic}, instead of choosing guaranteed close or distant pairs.

An established strategy to quantify social relations in the classroom is to form a \emph{friendship network} by asking students to list their friends, and using graph theory to analyze the collected data~\cite{george1996friendship, strauss2003social}. The particular form used in this study is a directed graph $G = (V,E)$, with participants as vertices $v\in V$, and nominations as directed edges $e\in E$. While the friendship network is a coarse approximation of the complex social relationships in the classroom, it allows us to introduce a \emph{social distance} heuristic $d(a,b)$, described in Algorithm \ref{alg:symmetric-distance}. 


\begin{algorithm}[h]
    \caption{\textit{Social distance heuristic}: mathematical distance measuring separation between individuals in a directed friendship network $G=(V, E)$. Individuals are represented by vertices $v\in V$; friendship nominations are represented by directed edges $e\in E$.}
    \label{alg:symmetric-distance}
    \begin{algorithmic}
        \Require Directed graph $G = (V, E)$. \Comment{Represents the friendship network.}
        \Require Vertices $a, b \in V$. \Comment{Represent individuals in the network.}
        \If{paths from $a$ to $b$ exist}
            \State $d'(a,b) \gets $ (length of the shortest path connecting $a$ to $b$).
        \Else
            \State $d'(a,b) \gets |V|.$ \Comment{Paths can be at most $|V|-1$ edges long.}
        \EndIf
        \State $d(a,b) \gets d'(a,b) + d'(b, a).$ \Comment{$2 \leq d(a,b) \leq 2|V|$, assuming $a\neq b$.}
    \end{algorithmic}
\end{algorithm}

Our solution to obtain balanced experimental conditions is to ask each child to participate twice; once in each condition. In the \textit{high-rapport} condition, we split the cohort into \textit{low-distance pairs} by \textit{minimizing} the sum of distances; in the \textit{low-rapport} condition, we split the cohort into \textit{high-distance pairs} by \textit{maximizing} the sum of distances. If we consider the fully connected weighted graph $\widetilde{G} = (V, \widetilde{E})$ having the participants as vertices, and having the (bidirectional) edge between $a$ and $b$ weighted by $d(a, b)$, the high-rapport partition corresponds to solving the \emph{minimum-weight maximal matching} problem; while the low-rapport condition corresponds to solving the \emph{maximum-weight maximal matching} problem.

\subsubsection*{Validation of the Pair-Making Procedure}
\label{sec:pair-validation}

In the weeks before performing the main study, we asked each participant to privately fill in a form listing their \emph{``closest friends''}. We refer to this document as the \textit{friendship nomination} form. The form contained the prompt and 10 empty lines, with no specific instruction on how many names to include. Respondents were instructed to identify at least one close friend, and encouraged to list a good amount of friends. While respondents were encouraged to list classmates from school, they were also allowed to add any other names.

All 38 included participants filled in the friendship nomination. After matching the nominations to IDs with help from the teaching team, we obtained $2.68 \pm 1.97$ within-cohort nominations per child (min 0, max 8). Since only one nomination involved students from different years, all subsequent analysis was performed separately per year.

\begin{figure}[!h]
    \centering
    \includegraphics[width=0.90\textwidth]{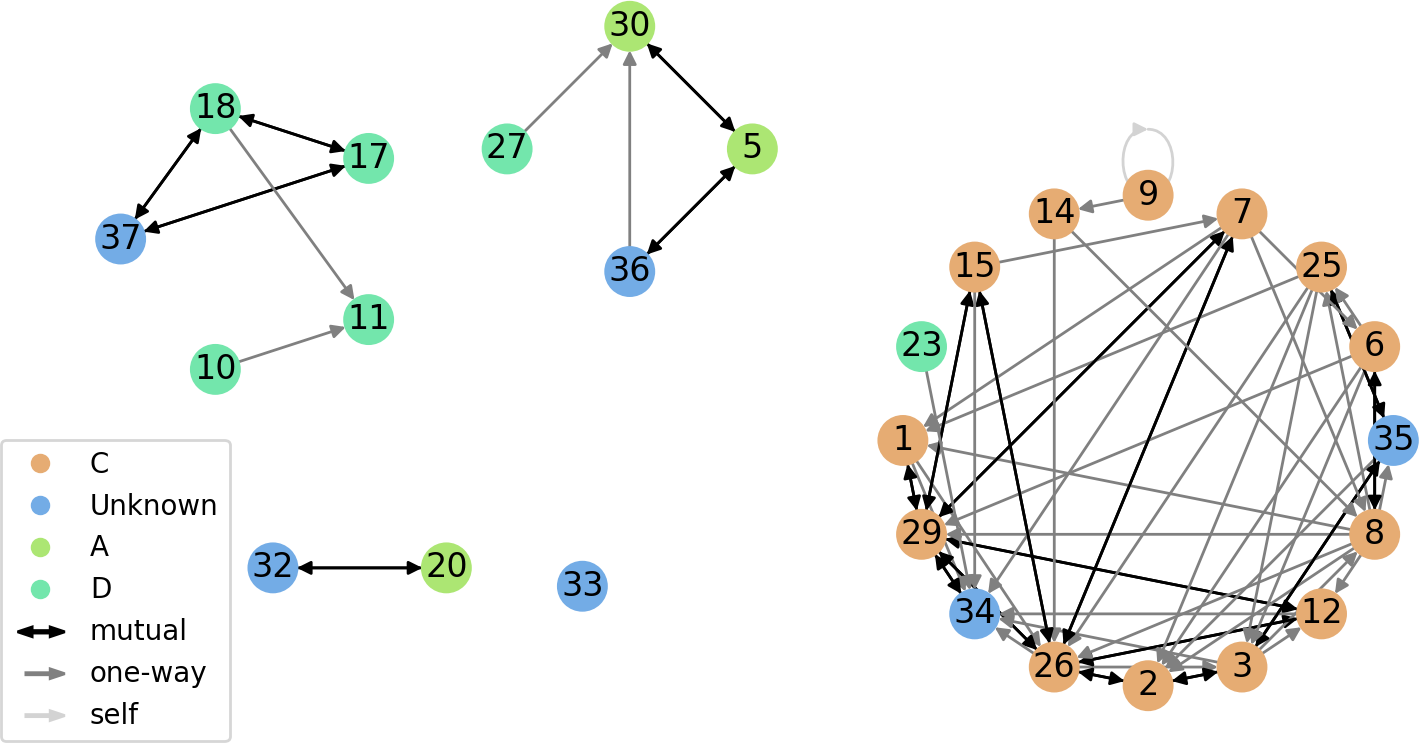}
    \caption{Friendship network of the Year 2 cohort. Vertex color indicates class (labeled randomly as A, C, D for anonymity). Edge color and shape indicate type of connection (one-way or mutual). The light gray loop corresponds to a child who nominated themselves.}
    \label{fig:network-y2}
\end{figure}

\begin{figure}[!h]
    \centering
    \includegraphics[width=0.70\textwidth]{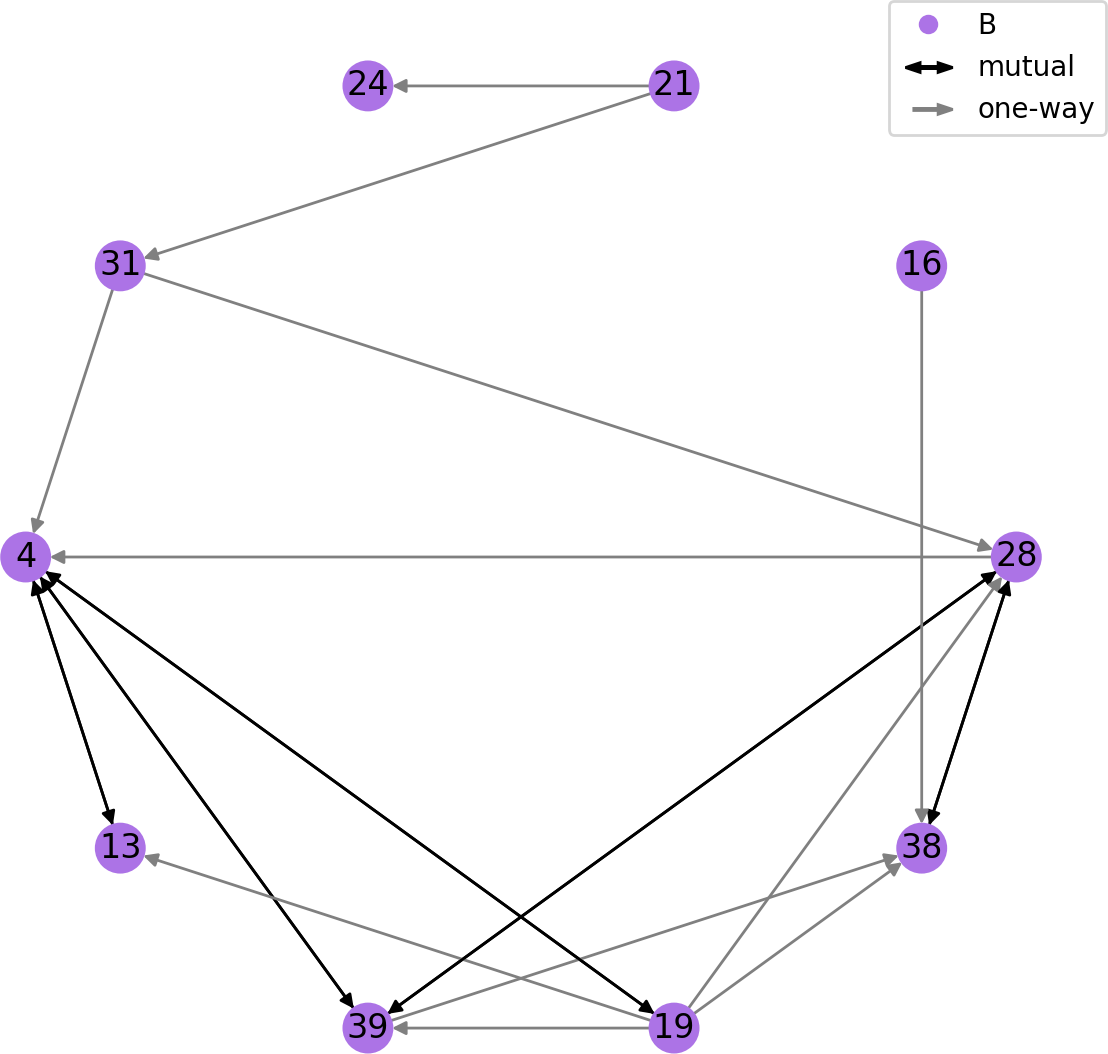}
    \caption{Friendship network of the Year 3 cohort. All children belonged to the same class (labeled randomly as B; shown in purple). Edge color and shape indicate type of connection (one-way or mutual).}
    \label{fig:network-y3}
\end{figure}

Figure \ref{fig:network-y2} shows the friendship network for Year 2. We can see the subgraphs corresponding to each class tend to be quite connected, while very few connections exist accross class boundaries. We hypothesize this helped the quality of the pair-making process. Figure \ref{fig:network-y3} shows the friendship network for Year 3. Since only one class participated, the graph is more evenly connected, although some students are still completely separated from each other.

We used our pair-making procedure to obtain two partitions of the student population (a high-rapport pairing and a low-rapport pairing), calculated separately per year. This resulted in 14 pairs per condition in Year 2, as well as 5 pairs per condition in Year 3, totalling 38 pairs (19 per condition). The pairings were validated by the teaching team, who confirmed no problematic pairs were suggested by the algorithm (e.g., children likely to get in a fight), and that the high-rapport pairings generally corresponded to closer acquaintances than the low-rapport pairings.

Table \ref{tab:partition-stats-distance} shows statistics for the social distance heuristic in each year and condition. In both cohorts, we obtain a desired lower social distance for the high-rapport condition, with large effect sizes (as measured by Cohen's d~\cite{cohen1988statistical}), but the effect is stronger in Year 2, and a t-test only indicates significant results in Year 2. Given the sample size, the lack of significance in Year 3 is expected; however, we are confident that the effect size supports our choice of methodology. Note that distances depend on the cohort size: in Year 2, distances are in range $2 \leq d(a,b) \leq 56$, while in Year 3, distances are in range $2 \leq d(a,b) \leq 20$.

\begin{table}[!ht]
    \caption{Statistics for the social distance heuristic. The p-value corresponds to a t-test with $\alpha=0.05$; asterisk denotes significance. The reported effect sizes (measured with Cohen's d) are typically described as large to huge \cite{sawilowsky2009new}. The statistics are calculated separately for each year. Note the small sample sizes.}
    \label{tab:partition-stats-distance}
    \begin{center}
        \begin{tabular}{|cl|r|rr|cc|}
            \hline
            \textbf{Year} & \textbf{Condition} & \textbf{count} & \textbf{mean} & \textbf{std} & \textbf{p (t-test)} & \textbf{Cohen's d} \\
            \thickhline
            \multirow{2}{*}{2} & high rapport & 14 & 13.64 & 17.40 & \multirow{2}{*}{$< 0.001^*$} & \multirow{2}{*}{3.08} \\
            & low rapport & 14 & 54.21 &  6.68 & & \\
            \hline
            \multirow{2}{*}{3} & high rapport & 5 &  7.60 &  4.67 & \multirow{2}{*}{0.12} & \multirow{2}{*}{1.07} \\
            & low rapport & 5 & 12.80 &  5.02 & & \\
            \hline
        \end{tabular}
    \end{center}
\end{table}

\subsection*{Task Design}
\label{sec:task-design}

The primary activity in this study is a collaborative storytelling game in which a pair of children develop and present a story together, with the help of a virtual deck of picture cards. In the \textit{planning phase}, the children are led into a private space (the \textit{planning area}), and a random selection of 12 virtual cards is presented face-down in a standing touchscreen. The children can reveal as many cards as they see fit by pressing on them, and are instructed to select a subset of 6 cards to be incorporated into the story. They are allowed as much time as needed to plan the story and to move freely within the planning area. Later, in the \textit{presentation phase}, the children jointly present the story they designed to the experimenters in a designated \textit{presentation area}. The planning and presentation phases are then repeated for each additional game round.

\begin{figure}[!h]
    \centering
    \includegraphics[width=\textwidth]{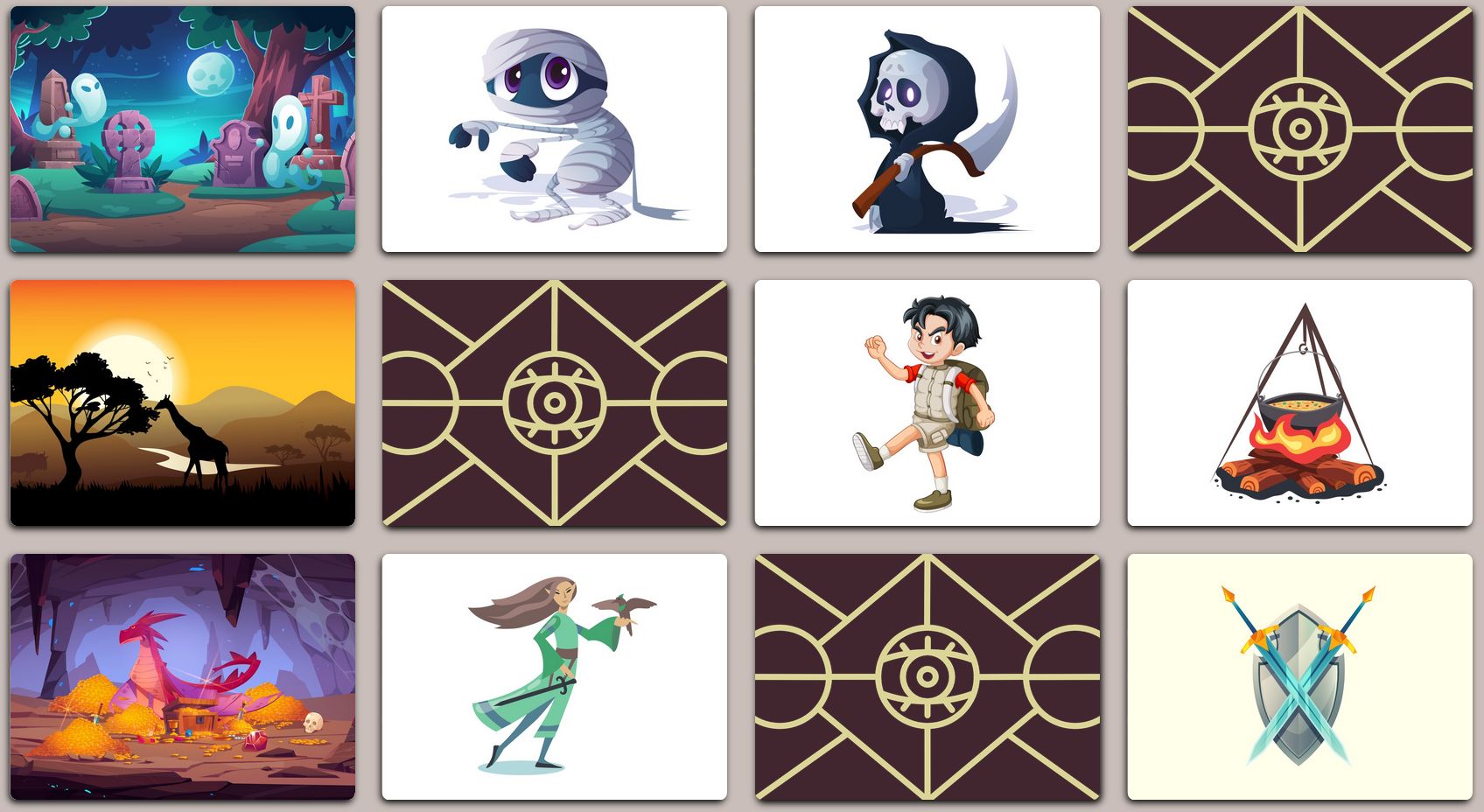}
    \caption{Example of a playing board, showing face-up and face-down cards. Each row corresponds to one setting. Top to bottom: halloween monsters, hiking, fantasy. Each column corresponds to a card type. Left to right: location, character, character, object.}
    \label{fig:board}
\end{figure}

The card game was implemented as a static webpage, using basic web technologies (HTML, JavaScript, CSS). The cards displayed to the children are chosen at random from a pool of \textit{settings} and \textit{types}. There are nine settings available, chosen to be a mix of realistic scenarios (e.g., hiking in nature, going to the hospital) and fantastic scenarios (e.g., knights and wizards, halloween monsters). All cards are displayed in a similar cartoon style, in order to promote free mixing of options. Each setting contains a number of location cards, character cards, and object cards. In each round of the game, three settings are chosen at random (ensuring the same option is not chosen in two consecutive round). One location, two characters, and one object are chosen from each scenario, forming a grid of 12 cards. We collected the original card images from the free-use image website Freepik\footnote{\url{https://www.freepik.com/}}, and processed them to a uniform size. Figure \ref{fig:board} shows an example board with most cards revealed.

\subsection*{Hypotheses and Measures}

In \emph{\nameref{sec:pair-validation}}, we provided statistical evidence that the pair-making procedure attained larger pair distances in the low-rapport condition when compared to the high-rapport condition, but we did not address whether rapport was successfully manipulated. In order to measure social and emotional effects, we employed questionnaires to validate the following experimental hypotheses:
\begin{itemize}
    \item[H1] Children feel closer to their partner when playing in the high-rapport condition than when playing in the low-rapport condition.
    \item[H2] Children report higher scores in the Valence-Arousal-Dominance model of emotion when playing in the high-rapport condition than when playing in the low-rapport condition.
\end{itemize}

\begin{figure}[!h]
    \centering
    \includegraphics[width=0.8\textwidth]{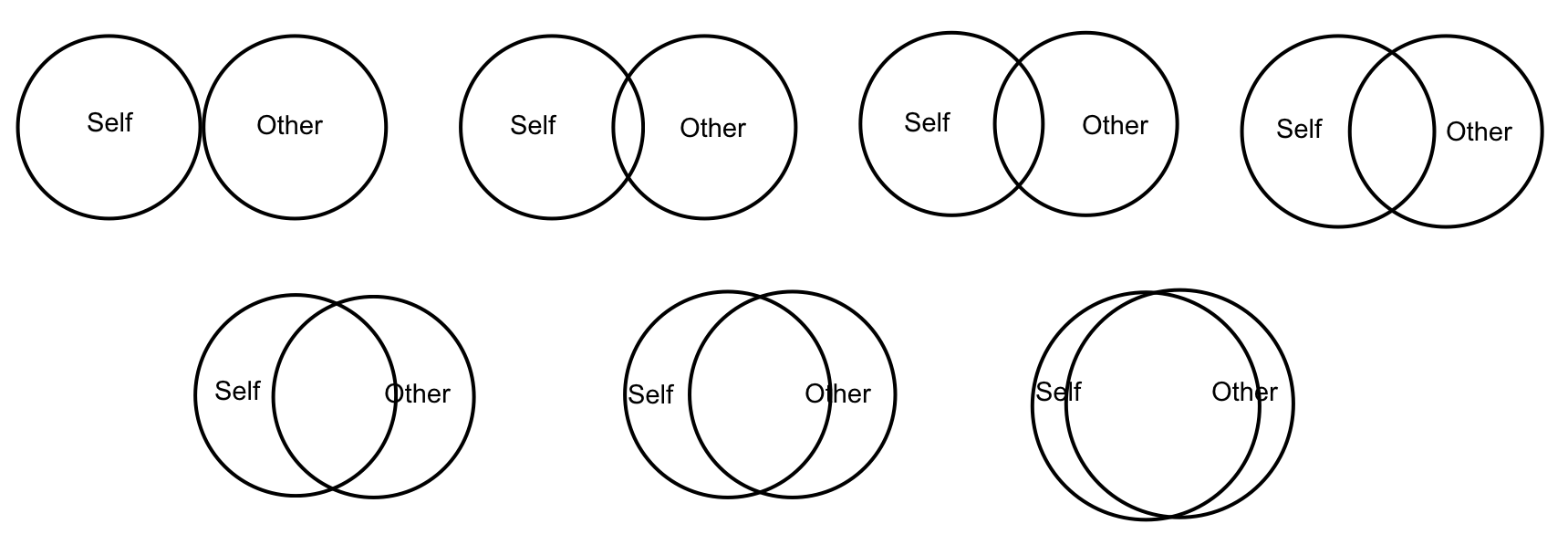}
    \caption{IOS questionnaire~\cite{aron1992inclusion}, used to measure the closeness of each participant with their best friend, a ``bad guy'' from fictional media, and their partner during the game. Measured after the interaction.}
    \label{fig:ios}
\end{figure}

We tested H1 through a post-interaction administration of \textit{Inclusion of Other in the Self} (IOS)~\cite{aron1992inclusion} (shown in Figure \ref{fig:ios}), a single-item 7-point pictographic scale used to measure a person's perceived closeness to another individual. Following Kory-Westlund~\cite{westlund2018measuring}, we add two calibration items to verify that the children understood the questionnaire. We asked the children to rate their closeness with three individuals: (1) their best friend, (2) a \textit{``bad guy''} from fictional media, and (3) their partner during the storytelling activity.

\begin{figure}[!h]
    \centering
    \includegraphics[width=0.6\textwidth]{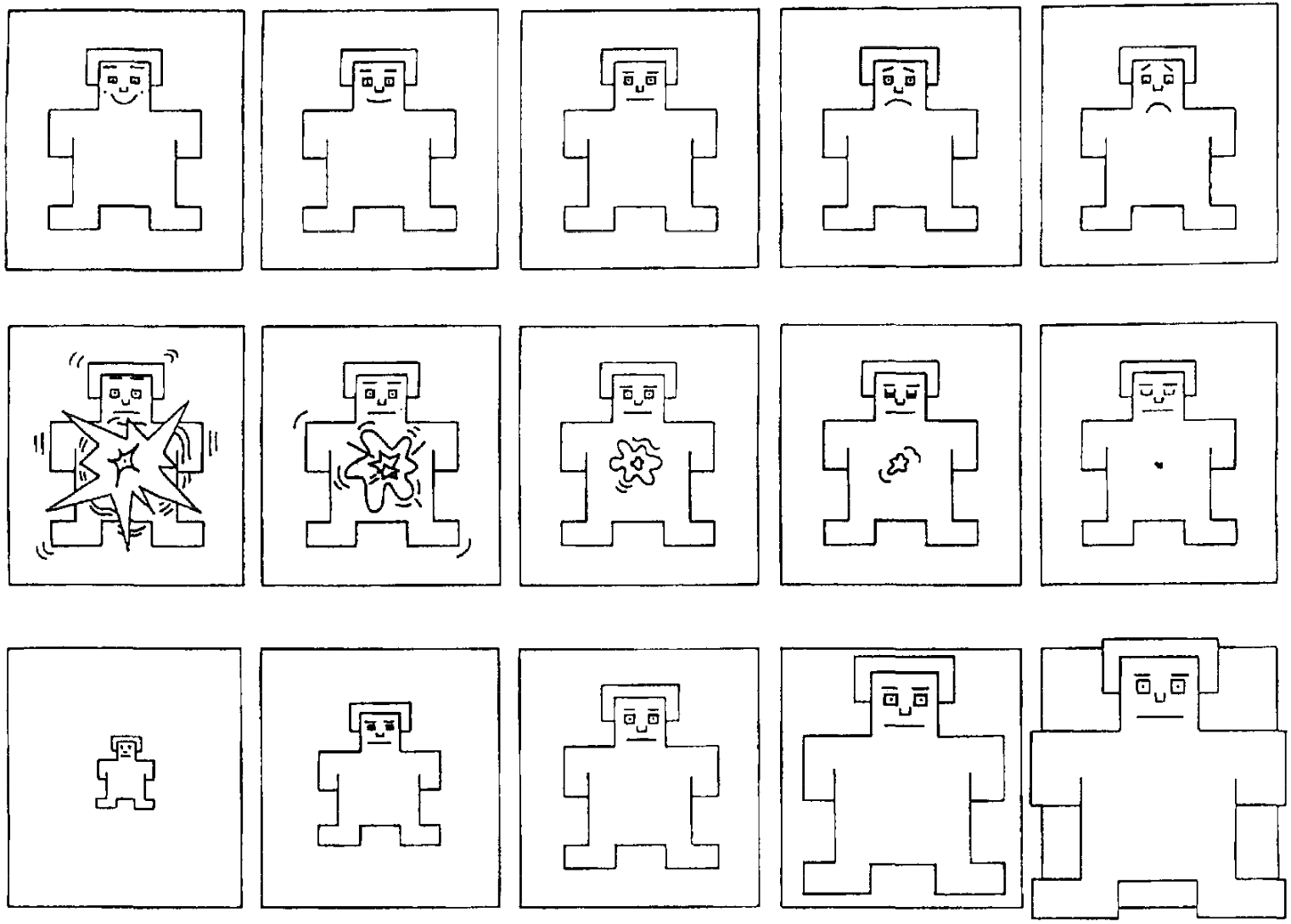}
    \caption{SAM questionnaire~\cite{bradley1994measuring}, used to measure the Valence (top row, left to right), Arousal (middle row, right to left), and Dominance (bottom row, left to right) dimensions of emotion. Measured before and after the interaction.}
    \label{fig:sam}
\end{figure}

Regarding H2, the Valence-Arousal-Dominance model is a continuous-variable representation of emotion~\cite{russell1980circumplex}. Valence represents pleasantness of emotion, Arousal measures intensity of emotion, and Dominance refers to the degree of control experienced. We tested H2 through pre-interaction and post-interaction administration of the \textit{Self-Assessment Manikin} (SAM)~\cite{bradley1994measuring} (shown in Figure \ref{fig:sam}), a pictographic questionnaire used to measure a person's emotional state. It consists of three single-item scales directly measuring Valence, Arousal, and Dominance. Each scale contains 5 pictures and is graded 1-5, but the original procedure allows adding crosses between pictures, making it a 9-point scale. Given our young audience, we chose to simplify it as a 5-point scale.

\subsection*{Experimental Apparatus}

\begin{figure}[!h]
    \centering
    \includegraphics[width= \textwidth]{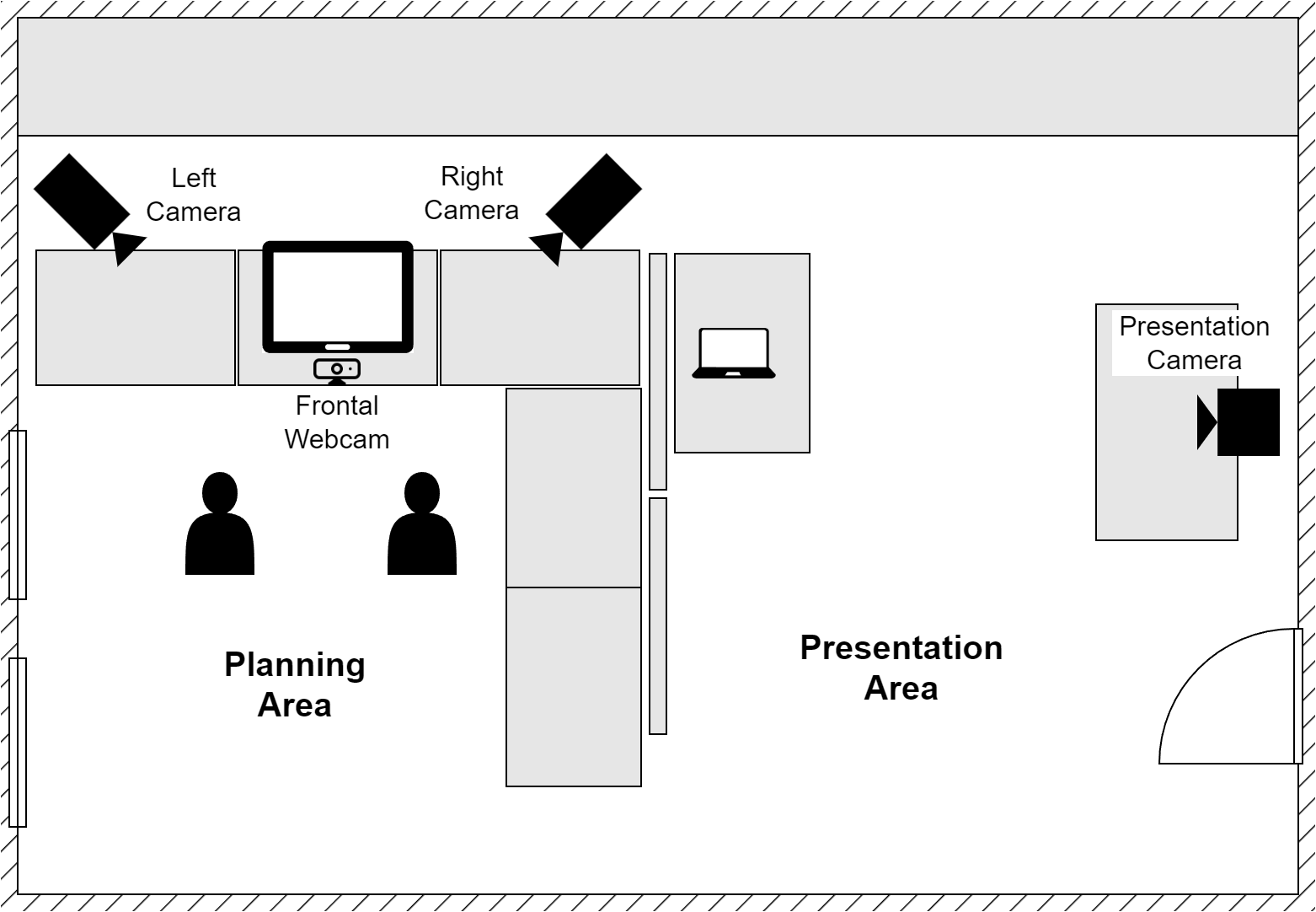}
    \caption{Diagram depicting the experimental area used for the storytelling activity. Children first designed a story in the \textit{planning area}, and then emerged into the \textit{presentation area} to retell the story to the experimenters.}
    \label{fig:setup}
\end{figure}

The main activity was performed inside the collaborating school, within the English library room. Figure \ref{fig:setup} illustrates the experimental setup.

The room's desks (wide rectangles in the diagram) were arranged to delineate the \textit{planning area}, a rectangular enclosure of approximately 160 cm $\times$ 140 cm where the participants performed the primary activity (planning a story). The \textit{planning area} was further separated from the rest of the room by a set of mobile walls (thin rectangles in the diagram), creating visual isolation from the area occupied by the researchers. 

One of the sides delimited by desks was designated as the front of the \textit{planning area}. A touchscreen was placed at its center in standing mode, and used to display the storytelling card game. Two Panasonic HC-V380 cameras ($1920 \times 1080$px at 25fps; stereo audio at 48kHz) were mounted on tripods at either side of the touchscreen, at a height of 180cm. These are referred from the child's point of view as \textit{left camera} and \textit{right camera}. A Logitech c920 webcam ($1920 \times 1080$px at 30fps; stereo audio at 48kHz) was placed right in front of the touchscreen to compensate for any occlusions caused by the screen itself. This is referred to as the \textit{frontal camera}.

Per-child audio of the whole interaction was recorded using head-mounted microphones. Each child was fitted with a SubZero SZW-20 wireless microphone set. The audio was captured with a Tascam US-2X2HR audio interface.

In contrast to the walled-off planning area, the \textit{presentation area} was the working space used by the experimenters. The \textit{presentation area} contained a Windows laptop used to (1) run the game application shown on the touchscreen, using the Firefox web browser; (2) record the frontal camera video stream, using Logitech recording software; and (3) record the headset audio streams, using Audacity (a popular open-source audio editing and recording application).

Once a participating pair had finished planning their story, they were instructed to enter the \textit{presentation area}, stand in front of the \textit{presentation camera}, and retell their story to the experimenters. A Canon EOS Rebel T3i ($1920 \times 1080$px at 29.97fps; stereo audio at 48kHz) was used in all sessions except one; an iPhone Pro 11 ($3840 \times 2160$px at 30fps; stereo audio at 44.1kHz) was used in the remaining session. The \textit{presentation camera} was mounted at a height of approximately 120 centimeters, 2 meters away from the children.

\subsection*{Procedure and Protocol}

The study started with an introductory meeting with each participating class. During the meeting, the whole classroom was introduced to the storytelling game, and allowed to try it in groups of 2-5 students. This was done to familiarize the children with the activity, and to register further interest from the students. During the following days, registered participants filled out the friendship nomination form in private. Each child was subsequently assigned to a high-rapport pair and a low-rapport pair, as indicated in \emph{\nameref{sec:pair-making}}.

For the remainder of the study, we invited one pair at a time to participate in the storytelling game. Pairs were chosen based on availability, prioritizing balance between experimental conditions to avoid order effects. Participating pairs were asked to communicate in English during the activity.

The pair were first called into the experimental space, where they played a short warm-up round of the game with help from two researchers, in order to remind them of the rules and familiarize them with the \textit{planning area} and \textit{presentation area}. Following this, the pair were separated, and each child filled the pre-activity questionnaire (SAM) with help from a researcher.

Next, the children donned the microphones with help from the researchers and proceeded to play several rounds of the game. They were allowed to play up to three rounds, or until the total planning time exceeded 10 minutes. Some exceptions were made for fast pairs, allowing a fourth round if the children requested it. The participants could ask to stop early at any time.

Finally, the children removed the microphones with help from the researchers and were separated again to fill the post-activity questionnaire (SAM and IOS).

\section*{Data Analysis}

\subsection*{Data Collection}
\label{sec:data-collection}

Of the 38 scheduled sessions, 33 were performed as planned. Two low-rapport sessions were canceled due to a lack of participant availability. One high-rapport session was discarded because it was a false positive: the name-based nomination system produced a mismatch due to two children sharing the same name. Finally, an error in communication caused a non-scheduled pair to play together, forcing us to substitute two distance-2 pairs (children had nominated each other directly) for two distance-4 pairs (children had nominated common friends). All children from the re-scheduled pairs were friends and could be found in the same playgroup. 

In total, 35 sessions were included in the recordings: 18 high-rapport sessions, totalling 57 rounds ($3.17 \pm 0.62$ rounds per session); and 17 low-rapport sessions, totalling 49 rounds ($2.88 \pm 0.70$ rounds per session). Proportionally, high-rapport data corresponds to 51.4\% of the \textit{sessions}, and 53.8\% of the \textit{rounds}.

For each of the 3 cameras recording the \textit{plannnig area} (left, right, frontal), as well as the 2 head-mounted microphones, we obtained one continuous recording covering all rounds of the game (except the warm-up round). Upon review, the frontal camera recordings were deemed low-quality and were discarded from further analysis. The main factors for this decision were a disproportionately high number of frames in which the participants' faces were out of view, and the fact that some children played with the camera, moving it around or even flipping it.

\subsection*{Data Processing}

The unprocessed audio sources (per-child headphone recordings) and video sources (left and right cameras) consisted of one continuous recording for each participating pair, containing all the game rounds that the pair played. Video sources were manually cut and processed using FFmpeg\footnote{\url{https://ffmpeg.org/}}; audio sources were similarly cut and processed within Audacity. A Python script was used to further synchronize the files based on audio content analysis. The package used for synchronization\footnote{\url{https://github.com/bbc/audio-offset-finder}} claims the \textit{``accuracy is typically to within about 0.01s''}; manual observation indicates no perceivable time differences.

After processing the data from all 35 included pairs, we obtained 106 sets of synchronized multimodal recordings, corresponding to the \textit{planning phase} of each game round played, for a total of 3h 40m of recorded interaction time. The \textit{presentation phase} recordings were not processed, and are left for future analysis. Table \ref{tab:durations} shows duration statistics, including per-round duration in each experimental condition. While mean comparison suggests high-rapport pairs played longer, standard deviations are high, and a cursory t-test did not indicate ordering of the variables.

\begin{table}[!ht]
    \caption{Duration of the processed recordings. Times per round are given as mean and standard deviation.}
    \label{tab:durations}
    \begin{center}
        \begin{tabular}{|r|ccc|}
            \hline
            & \textbf{high-rapport} & \textbf{low-rapport} & \textbf{overall} \\
            \thickhline
            \textbf{per-round} & 2m 03s $\pm$ 1m 08s & 2m 06s $\pm$ 1m 31s & 2m 04s $\pm$ 1m 19s \\
            \textbf{cumulative} & 1h 57m 01s & 1h 43m 06s & 3h 40m 07s \\
            \hline
        \end{tabular}
    \end{center}
\end{table}

\subsection*{Questionnaire Analysis}

From the collected questionnaire data, two entries were discarded due to unclear responses (the child selected multiple options, and the accompanying researcher could not clarify the child's intent). Further entries were discarded if they could not be paired within-subjects (because the child only participated in one accepted pair, or because the child had missing data in one of their questionnaire responses). 

After curation, the questionnaire data contained full sets of responses for 30 children, across 34 pairs. 17 children participated in the high-rapport condition first, and 13 children participated in the low-rapport condition first (56.67\% high-rapport first).

We used the Python package Pingouin\footnote{\url{https://pingouin-stats.org}}~\cite{vallat2018pingouin} to perform statistical analysis of the questionnaire data. Shapiro-Wilks tests suggest none of the questionnaire item responses are normally distributed, either overall or per condition ($p < 0.01$). Therefore, all subsequent analysis is done using nonparametric approaches. All statistical tests use the standard significance level $\alpha=0.05$.

\begin{figure}[!h]
   \centering
   \includegraphics[width=0.75\linewidth]{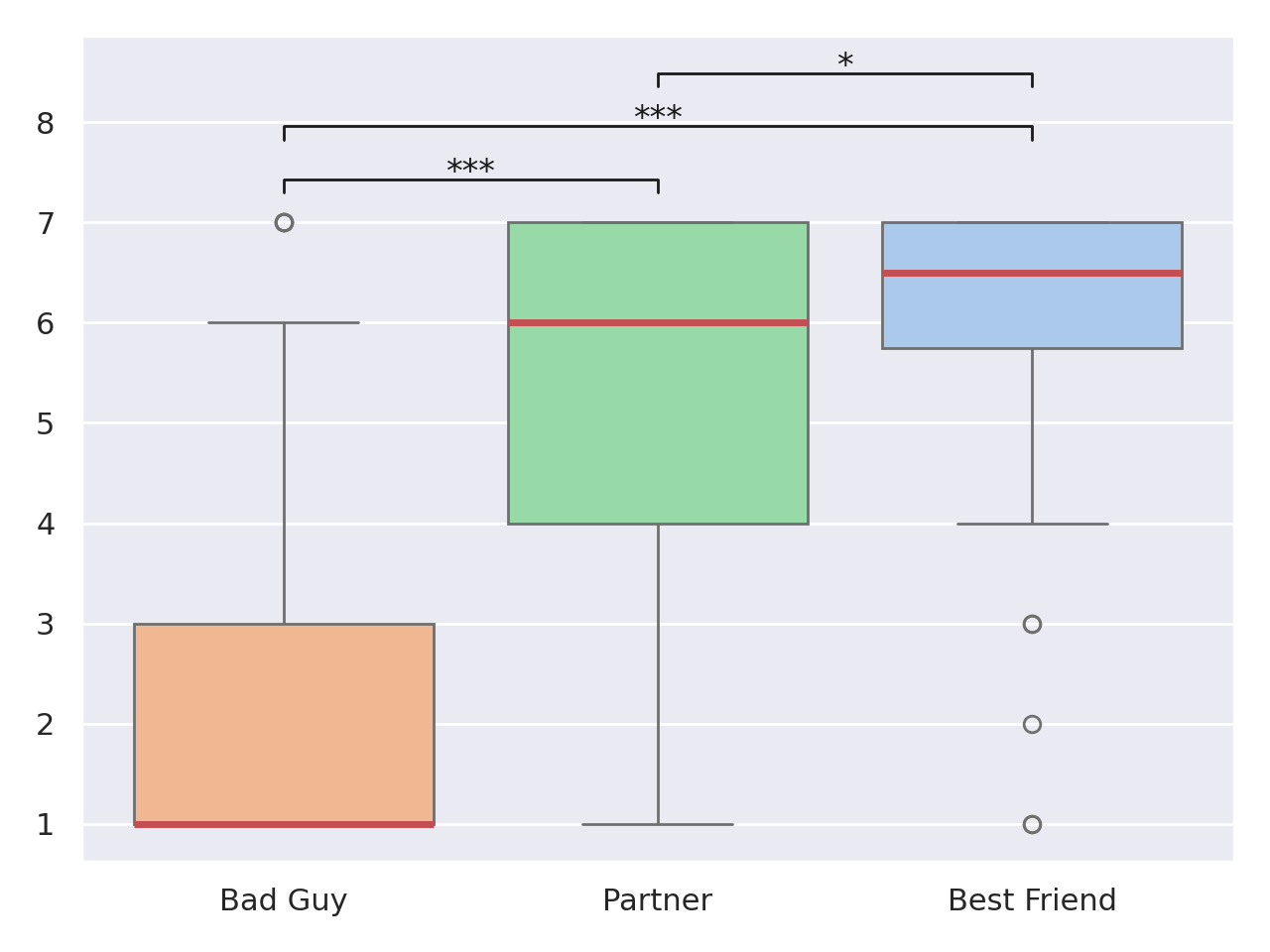}
   \caption{Response distribution for the IOS questionnaire items \emph{Bad Guy}, \emph{Partner}, and \emph{Best Friend}. Boxes show the 25\% (bottom), 50\% (red, middle), and 75\% (top) quartiles. Whiskers extend to the lowest and highest samples within 1.5 interquartile ranges from the box. Samples outside the whiskers are shown as circles. Significant differences are shown ($*p < 0.05, **p < 0.01, ***p < 0.001$).}
   \label{fig:stats-ios-overall}
\end{figure}

\begin{figure}[!h]
   \centering
   \includegraphics[width=0.75\linewidth]{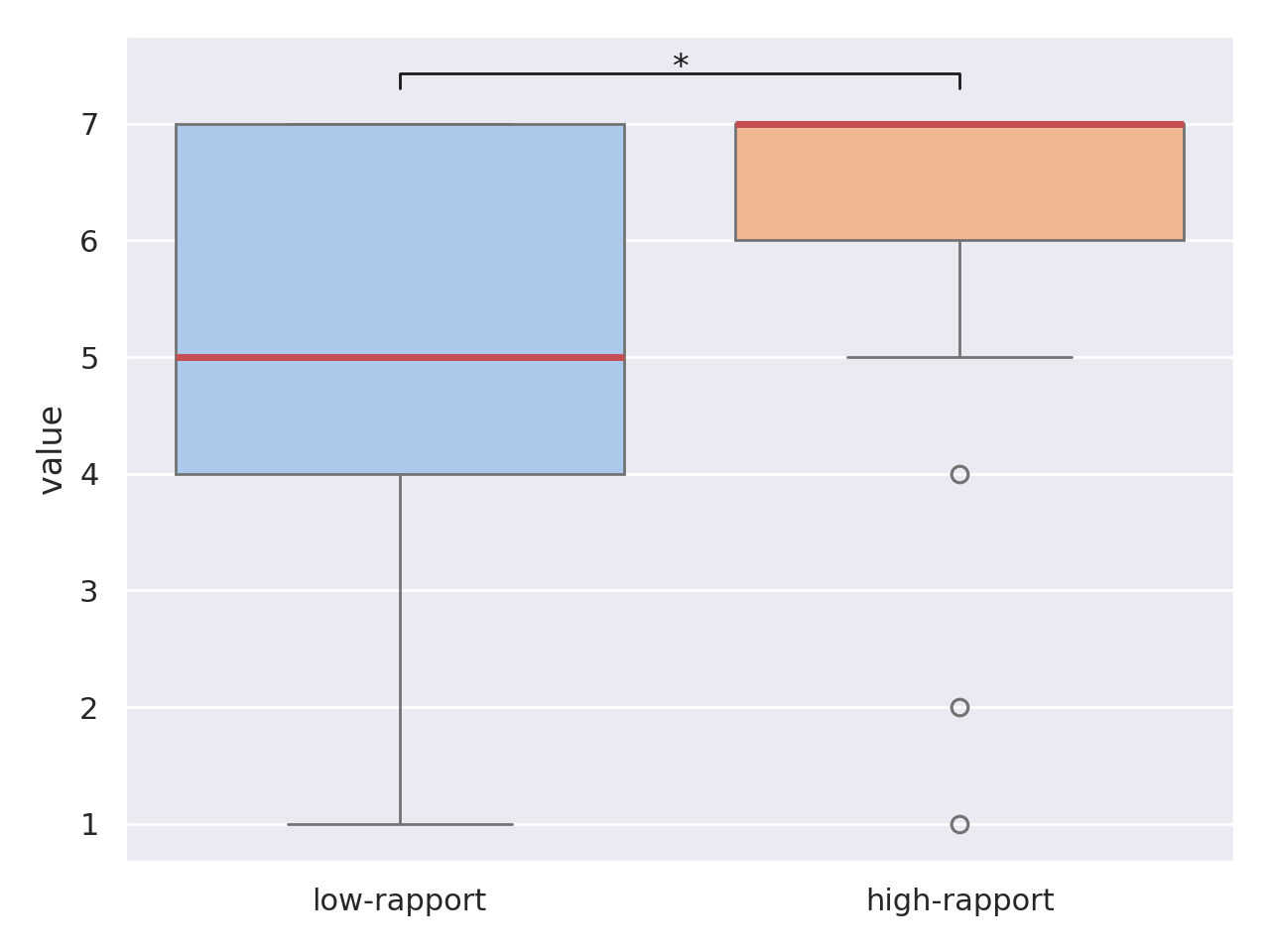}
   \caption{Response distribution for the IOS questionnaire item \emph{Partner}, under the high-rapport (orange) and low-rapport (blue) conditions. The difference is significant ($*p < 0.05$).}
   \label{fig:stats-ios-partner-condition}
\end{figure}

Overall response distributions for the IOS post-test are displayed in Figure \ref{fig:stats-ios-overall} as a boxplot, with significant differences highlighted. The control items \emph{Bad Guy} and \emph{Best Friend} followed the expected ordering with respect to the target item \emph{Partner}: A two-sided Wilcoxon signed-rank test showed that \emph{Bad Guy} scores are significantly lower than \emph{Partner} scores ($W=118, p < 0.001$), \emph{Partner} scores are significantly lower than \emph{Best Friend} scores ($W=221.5, p < 0.05$), and \emph{Bad Guy} scores are significantly lower than \emph{Best Friend} scores ($W=57$, $p < 0.001$). These results show that the children understood the IOS test. Figure \ref{fig:stats-ios-partner-condition} similarly shows a boxplot of the target item \emph{Partner} across experimental conditions. Another two-sided Wilcoxon test showed the scores in the high-rapport condition are significantly higher than in the low-rapport condition ($W=42.5, p<0.05$). This validates that the pairing scheme produced significantly closer pairs in the high-rapport condition, when compared to the low-rapport condition. Based on the aforementioned comparisons, we can conclude there is clear statistical evidence in favor of H1.

\begin{figure}[!h]
   \centering
   \includegraphics[width=0.75\linewidth]{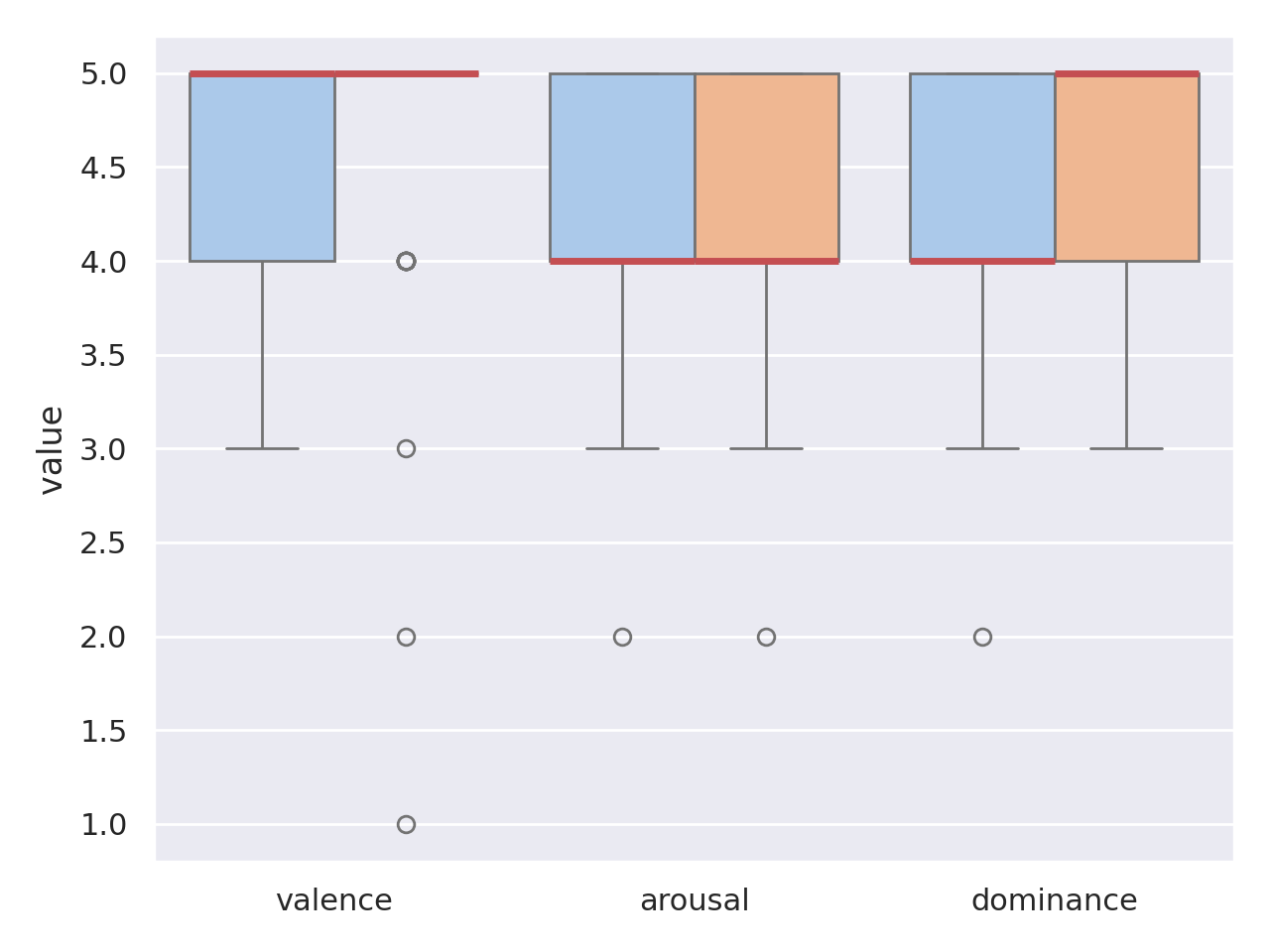}
   \caption{Response distribution for the SAM questionnaire items \emph{Valence}, \emph{Arousal}, and \emph{Dominance}, in the \emph{pre-test} (blue) and \emph{post-test} (orange). No significant relations were found.}
   \label{fig:stats-sam}
\end{figure}

Figure \ref{fig:stats-sam} shows the response distribution for each SAM item, comparing pre-test to post-test responses. We observed a strong ceiling effect, causing the data to be inconclusive. Most participants immediately chose the options they perceived as most positive, resulting in a response distribution heavily biased towards the high end of each scale. In concordance with this observation, a cursory repeated-measures ANOVA analysis shows no significant effects of condition nor moment (pre-interaction vs. post-interaction) on the SAM responses. We are therefore forced to reject H2 due to a lack of information.

The observed results (strong support for H1 despite the small sample size; rejection of H2 due to a ceiling effect) suggest that our pair-making strategy successfully promoted the formation of high-rapport vs. low-rapport pairs at the population level.

\section*{The UpStory Dataset}

The UpStory dataset is publicly available at \url{https://zenodo.org/doi/10.5281/zenodo.12635620}. It is an anonymized feature-based dataset, containing OpenPose and OpenFace features extracted per-frame. Data extracted from both the \textit{left camera} and \textit{right camera} sources is available. It provides data for all 35 pairs included in the recordings, totalling 3h 40m 07s of interaction data sampled at 25Hz. The following per-pair metadata is included: experimental condition (high-rapport or low-rapport), social distance heuristic, school year, number of rounds played, and each child's ID. As observed in \emph{\nameref{sec:data-collection}}, this corresponds to a total of 106 game rounds (57 high-rapport rounds and 49 low-rapport rounds). The remainder of this section gives detailed information on the choice of feature extractors, the extracted features, and custom post-processing required to track individuals over time.

In recent years, OpenPose and OpenFace have become the reference feature extractors for pose estimation and facial feature estimation in the literature \cite{wu2020automatic, srivastava2020recognizing, alsofyani2021attachment}. For this reason, we chose these two feature extractors to obtain an anonymized version of the dataset that is amenable to ML research. When applied to our synchronized video sources (left and right cameras), they produce time series of features: an entry is produced for each tracked feature, for each frame in the recording, for each detected individual. In our case, we can obtain a 25Hz time series per child for each tracked feature. OpenPose provides the position and confidence of 25 body keypoints. OpenFace provides 2D and 3D gaze direction estimates; 2D and 3D keypoint locations for the eyes; 2D and 3D keypoint locations for the whole face; position and angle estimates for the face as a whole; intensity estimates for 17 AUs; and presence estimates for 18 AUs. AUs represent specific muscle activations that produce facial appearance changes~\cite{ekman1978facial}. Table \ref{tab:extracted-features} shows all extracted metadata and features, and gives a brief description of each item.

\begin{table}[!ht]
\begin{adjustwidth}{-2.25in}{0in} 
    \caption{Metadata and feature sets contained in the UpStory feature dataset. Metadata is given per-pair; features are provided as time series, captured at 25Hz.}
    \label{tab:extracted-features}
    \begin{center}
        \begin{tabular}{|lll|}
            \hline
            \textbf{origin} & \textbf{variable} & \textbf{details} \\
            \thickhline
            \textbf{metadata} & condition & Experimental condition (high-rapport or low-rapport). \\
                     & social distance & Symmetric distance in the friendship network. \\
                     & year & academic year the children belonged to. \\
                     & rounds & Total number of game rounds played by this pair. \\
                     & Child IDs & The IDs of the two participating children. \\
            \hline
            \textbf{OpenPose} & joint position & 25 body keypoints; (x,y) in pixels. \\
                     & joint confidence & 25 body keypoints; confidence as fraction. \\
            \hline
            \textbf{OpenFace} & timestamp & Time from beginning of recording, in seconds. \\
                     & confidence & Face detection confidence, as a fraction. \\
                     & success & Is this a successful registration? (0 or 1). \\
                     & gaze & Gaze vector as (x,y,z) components, per-eye. \\
                     & gaze angle & Overall gaze direction, as (x,y) angles. \\
                     & 2D eye landmark & 56 eye keypoints; (x,y) in pixels. \\
                     & 3D eye landmark & 56 eye keypoints; (X,Y,Z) in estimated millimeters. \\
                     & head position & head position as (x,y,z) in estimated millimeters. \\
                     & head rotation & head rotation as (x,y,z) angles. \\
                     & 2D face landmark & 68 face keypoints; (x,y) in pixels. \\
                     & 3D face landmark & 68 face keypoints; (X,Y,Z) in estimated millimeters. \\
                     & AU presence & 18 AU activations, as binary variables (0 or 1). \\
                     & AU intensity & 17 AU activations, as continuous values (0 to 5). \\
            \hline
        \end{tabular}
    \end{center}
\end{adjustwidth}
\end{table}

A processing challenge in this study is the fact that both participants could move freely in the experiment space. This means the feature extraction pipeline needs to deal with movement over time, occlusions, and temporary loss of detection. Neither OpenFace not OpenPose offer identification functionality, meaning their output needs to be processed to maintain stable child identity over time. This was achieved with a custom post-processing pipeline based on three principles: minimizing the distance each body keypoint moves between consecutive frames (accounting for reported detection confidence), caching the last known location when tracking is lost, and ensuring face and body identities are consistent with respect to each other. 

\begin{figure}[!h]
   \centering
   \includegraphics[width=0.75\linewidth]{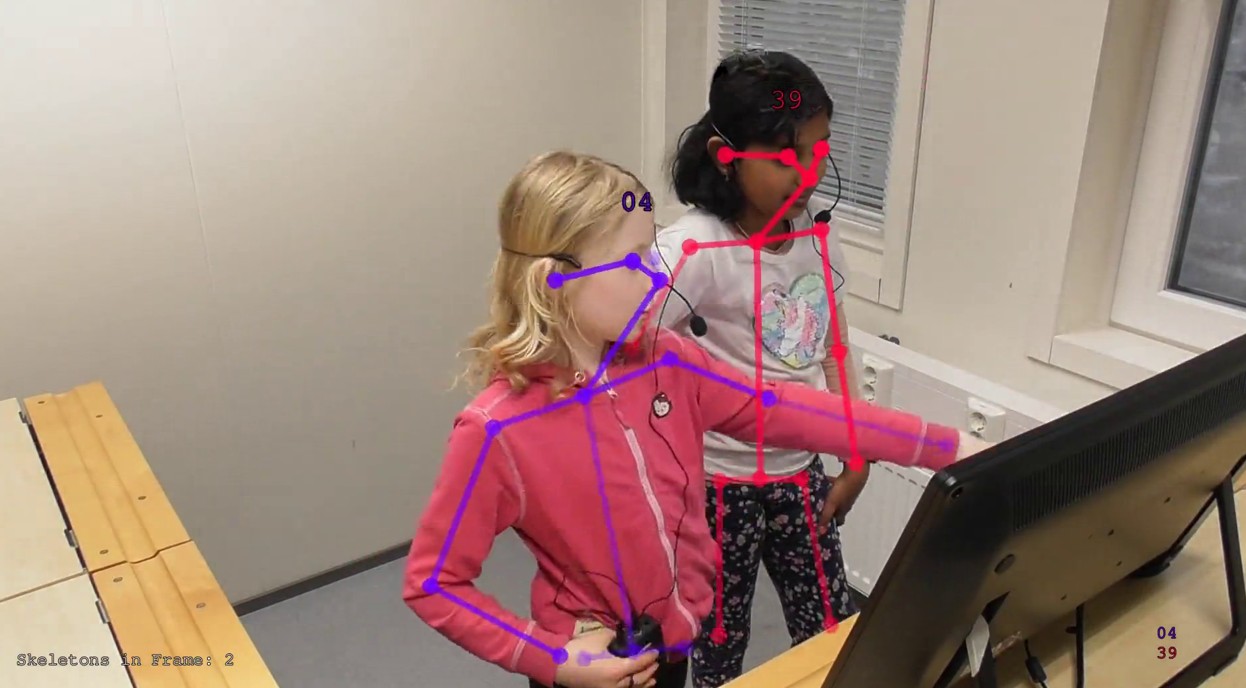}
   \caption{OpenPose output visualization, displaying the joint positions and confidences, overlayed on the original \textit{right camera} frame. Hue indicates child identity; transparency indicates detection confidence. In this frame, OpenPose successfully disambiguated a complex crossing of limbs. Child identity over time is not provided by either feature extraction tool, and is calculated as a post-processing step.}
   \label{fig:pose-viz}
\end{figure}

\begin{figure}[!h]
   \centering
   \includegraphics[width=0.75\linewidth]{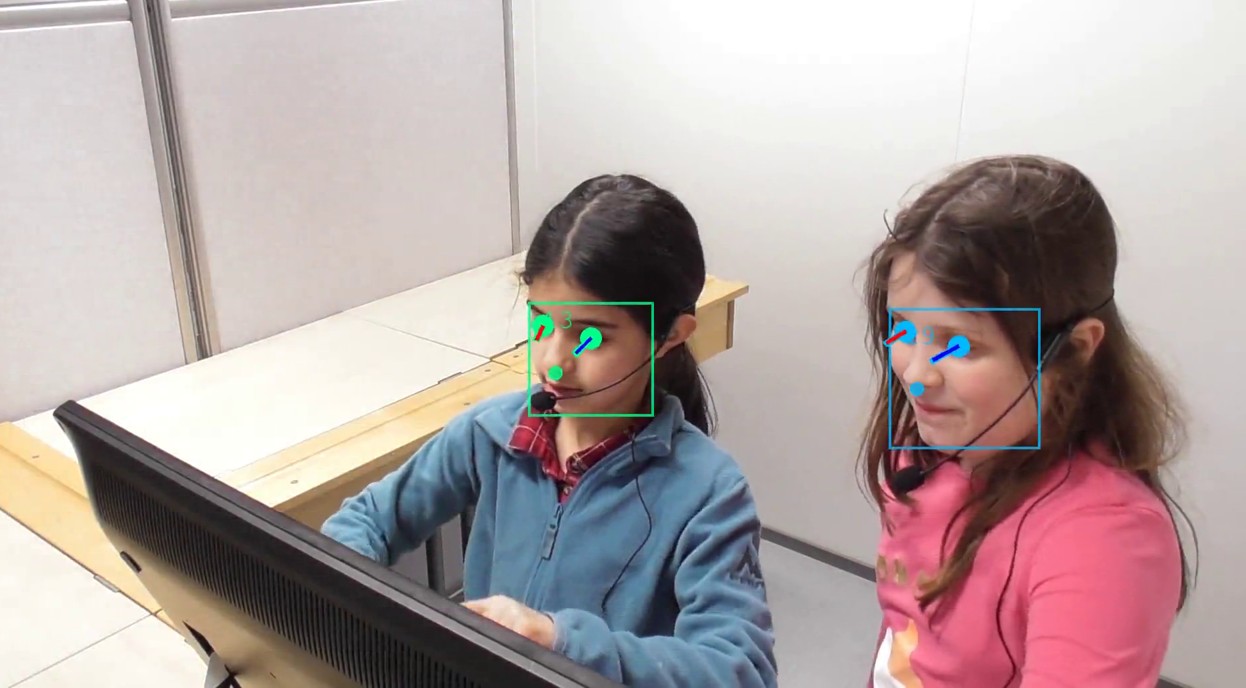}
   \caption{OpenFace output visualization, displaying the eye locations, gaze vectors, and face keypoint bounding box, overlayed on the original \textit{left camera} frame. Hue indicates child identity; gaze vectors are color-coded Blue for Left and Red for Right. Child identity over time is deduced from the corresponding body pose estimate.}
   \label{fig:face-viz}
\end{figure}

Figure \ref{fig:pose-viz} shows post-processed OpenPose output with attached identities (represented by the skeleton color and the ID tag). Each detected body keypoint is shown as a circle, with opacity indicating detection confidence. Similarly, Figure \ref{fig:face-viz} shows post-processed OpenFace output with attached identities. For simplicity, we only display the eye positions, gaze vectors, and bounding boxes containing all face keypoints.

\section*{ML Baselines}

In order to validate the predictive power of the UpStory dataset, we established baselines for the prediction of the level of rapport based on the publicly available features. Two different baselines are provided: \emph{\nameref{sec:ml:single-child}} considers data from a single child as a sample, while \emph{\nameref{sec:ml:pair}} considers joint data from both children in a pair as a sample. In both cases, the task is binary classification: using the AU feature time-series as input, the algorithm predicts the experimental condition (high-rapport pair vs. low-rapport pair). We first discuss \emph{\nameref{sec:ml:feature-selection}}, \emph{\nameref{sec:ml:data-stratification}}, and \emph{\nameref{sec:ml:model-selection}}, since they apply to both experiments.

\subsection*{Feature Selection}
\label{sec:ml:feature-selection}

Following an established procedure~\cite{paetzel2017investigating, srivastava2020recognizing}, we focused on the AU estimates provided by OpenFace, and reduced each AU time series to a selection of summary statistics, calculated per-child. As listed in Table \ref{tab:extracted-features}, OpenFace produces independently calculated estimates of \textit{presence} (binary variable) and \textit{intensity} (continuous variable); 17 AUs are covered in both modalities. For each AU estimate, we calculated the following summary statistics: mean, standard deviation, and 95\% percentile. The mean and standard deviation have been used by Paetzel~\cite{paetzel2017investigating} and by Srivastava~\cite{srivastava2020recognizing}, among others. Usage of the 95\% percentile is adapted from a similar idea in Alsofyani and Vinciarelli's work~\cite{alsofyani2021attachment}. If we consider all possible combinations of AU, estimate type, and summary statistic, we obtain 102 features (e.g., AU10-intensity-q95, or AU17-presence-mean). Table \ref{tab:considered-features} shows the possible combinations.

\begin{table}[!ht]
    \caption{A total of 102 features are considered for the ML baselines. Each feature is a combination of (1) a target AU, (2) an estimate type, and (3) a summary statistic. E.g., AU06-presence-mean (we track the AU06 presence estimate, and aggregate by taking the mean), or AU12-intensity-q95 (we track the AU12 intensity estimate, and aggregate by taking the 95\% quantile). This table lists all considered values for (1), (2), and (3). 
    In particular, only AUs with both estimates available are considered.
    q95 indicates 95\% quantile; std indicates standard deviation.}
    \label{tab:considered-features}
    \begin{center}
        \begin{tabular}{|lll|}
            \hline
            \textbf{AU} & \textbf{estimate type} & \textbf{statistic} \\
            \thickhline
            AU01, AU02, AU04, AU05, AU06, AU07, & presence & mean \\
            AU09, AU10, AU12, AU14, AU15, AU17, & intensity & std \\
            AU20, AU23, AU25, AU26, AU45 & & q95 \\
            \hline
        \end{tabular}
    \end{center}
\end{table}

While other authors working with larger datasets have chosen to use all 17 AUs as input features~\cite{alsofyani2021attachment}, due to our comparatively small number of samples, we decided to train on small feature sets consisting of 1-4 features to avoid overfitting. Two different feature selection approaches were combined: theory-based and data-driven. 

The theory-based feature sets are based on expressed happiness: the joint activation of AU06 (cheek raiser) and AU12 (lip corner puller) is identified by Ekman as the indicator of a \emph{``genuine smile''}, and associated with the basic emotion of happiness~\cite{ekman1978facial}. We considered both presence and intensity estimates (without mixing estimate types). As summary statistics, we either took the more established combination of mean and standard deviation, or alternatively the 95\% quantile for extreme event quantification. In total, 4 theory-based feature sets were considered. They are listed as the last block in Table \ref{tab:ml:feature-sets}.

For the data-driven approach, we used a battery of t-tests with Bonferroni correction to find variables having significantly different values between experimental conditions. We chose combinations of significantly different variables based on the following soundness rules: do not use more than 2 AUs at once, only use one type of statistic at a time (with mean and standard deviation being allowed as either two separate statistics, or a double combination), do not mix presence and intensity variables.

Table \ref{tab:ml:significant-features-left} lists the features that were significantly different across experimental conditions, as extracted from the \textit{left camera}, and provides the t-test results (with Bonferroni correction). Table \ref{tab:ml:significant-features-right} lists the corresponding features and statistics for the \textit{right camera}. In both sources, AU12 (lip corner puller) showed significant differences in standard deviation and 95\% quantile for both estimate types, though the difference in means was not significantly different under Bonferroni correction. AU06 (cheek raiser), AU10 (upper lip raiser) and AU25 (lips part) displayed significant differences in the left camera, while AU26 (jaw drop) displayed significant differences in the right camera. Note that AU25 and AU26 can indicate speech, suggesting that the amount of speech might act as a predictor of the experimental condition. Notably, only one of the significant statistics is a mean. This suggests children in the high-rapport condition expressed a wider range of emotions, with more extreme values being detected.

\begin{table}[!ht]
    \caption{Significantly different summary AU statistics accross experimental conditions for the \textit{left camera}, as measured by a t-test with Bonferroni correction. q95 indicates 95\% quantile; std indicates standard deviation.}
    \label{tab:ml:significant-features-left}
    \begin{center}
        \begin{tabular}{|lll|rr|}
            \hline
            \textbf{AU} & \textbf{estimate type} & \textbf{statistic} & \textbf{adjusted p-value} & \textbf{Cohen's d} \\
            \thickhline
            AU10 & presence  & std &    0.006  & 0.58 \\
                 &           & q95 & $< 0.001$ & 0.70 \\
                 & intensity & q95 &    0.012  & 0.55 \\
            \hline
            AU12 & presence  & std &    0.015  & 0.54 \\
                 &           & q95 &    0.012  & 0.55 \\
                 & intensity & std &    0.006  & 0.57 \\
                 &           & q95 &    0.010  & 0.55 \\
            \hline
            AU25 & presence  & std &    0.024  & 0.53 \\
                 & intensity & std &    0.026  & 0.52 \\
                 &           & q95 &    0.038  & 0.51 \\
            \hline
            AU06 & intensity & std &    0.020  & 0.52 \\
                 &           & q95 &    0.031  & 0.51 \\
            \hline
        \end{tabular}
    \end{center}
\end{table}

\begin{table}[!ht]
    \caption{Significantly different summary AU statistics across experimental conditions for the \textit{right camera}, as measured by a t-test with Bonferroni correction. q95 indicates 95\% quantile; std indicates standard deviation.}
    \label{tab:ml:significant-features-right}
    \begin{center}
        \begin{tabular}{|lll|rr|}
            \hline
            \textbf{AU} & \textbf{estimate type} & \textbf{statistic} & \textbf{adjusted p-value} & \textbf{Cohen's d} \\
            \thickhline
            AU12 &  presence &  std &    0.016  & 0.55 \\
                 &           &  q95 &    0.017  & 0.56 \\
                 & intensity &  std &    0.009  & 0.56 \\
                 &           &  q95 &    0.005  & 0.58 \\
            \hline
            AU26 &  presence & mean &    0.001  & 0.62 \\
                 &           &  std & $< 0.001$ & 0.72 \\
            \hline
        \end{tabular}
    \end{center}
\end{table}

Each AU selected by the data-driven approach was considered on its own, and combined with other AUs that were significant in the same source (e.g., AU10 and AU12 --- left camera ---, but not AU25 and AU26 --- mix of cameras). For each of the selected AU combinations, 95\% quantiles of the presence and intensity estimates were considered. Finally, the mean and standard deviation of the presence estimates was also considered. In total, 21 data-driven feature sets were considered. Together with the theory-driven feature sets, a total of 25 feature sets were considered. All combinations are listed in Table \ref{tab:ml:feature-sets}. The first three blocks correspond to features identified from the left camera; the third block corresponds to features identified from the right camera.

\begin{table}[!ht]
\begin{adjustwidth}{-2.25in}{0in} 
    \caption{Feature sets considered for ML training. Each row shows a feature set consisting of 1-4 individual features. Each feature represents a choice of tracked AU, estimate type, and summary statistic, as shown in Table \ref{tab:considered-features}. AU06 and AU12 combinations (last block) are chosen based on prior literature; other combinations are surfaced through data analysis.}
    \label{tab:ml:feature-sets}
    \begin{center}
        \begin{tabular}{|llll|}
            \hline
            AU10-presence-q95   & AU12-presence-q95  &                     &                    \\
            AU10-intensity-q95  & AU12-intensity-q95 &                     &                    \\
            AU10-presence-q95   &                    &                     &                    \\
            AU10-intensity-q95  &                    &                     &                    \\
            AU12-presence-q95   &                    &                     &                    \\
            AU12-intensity-q95  &                    &                     &                    \\
            AU10-presence-mean  & AU10-presence-std  & AU12-presence-mean  & AU12-presence-std  \\
            AU10-presence-mean  & AU10-presence-std  &                     &                    \\
            AU12-presence-mean  & AU12-presence-std  &                     &                    \\
            \hline
            AU25-presence-q95   & AU12-presence-q95  &                     &                    \\
            AU25-intensity-q95  & AU12-intensity-q95 &                     &                    \\
            AU25-presence-q95   &                    &                     &                    \\
            AU25-intensity-q95  &                    &                     &                    \\
            AU25-presence-mean  & AU25-presence-std  & AU12-presence-mean  & AU12-presence-std  \\
            AU25-presence-mean  & AU25-presence-std  &                     &                    \\
            \hline
            AU26-presence-q95   & AU12-presence-q95  &                     &                    \\
            AU26-intensity-q95  & AU12-intensity-q95 &                     &                    \\
            AU26-presence-q95   &                    &                     &                    \\
            AU26-intensity-q95  &                    &                     &                    \\
            AU26-presence-mean  & AU26-presence-std  & AU12-presence-mean  & AU12-presence-std  \\
            AU26-presence-mean  & AU26-presence-std  &                     &                    \\
            \hline
            AU06-presence-q95   & AU12-presence-q95  &                     &                    \\
            AU06-intensity-q95  & AU12-intensity-q95 &                     &                    \\
            AU06-presence-mean  & AU06-presence-std  & AU12-presence-mean  & AU12-presence-std  \\
            AU06-intensity-mean & AU06-intensity-std & AU12-intensity-mean & AU12-intensity-std \\

            \hline
        \end{tabular}
    \end{center}
\end{adjustwidth}
\end{table}

\subsection*{Data Stratification}
\label{sec:ml:data-stratification}

In order to provide high-quality validation and test sets, we partitioned the 
dataset into 5 \textit{folds} (disjoint subsets of samples), separated at the pair level, and using stratification to ensure good sample balancing. Sampling was designed to always distribute pairs from the same year and condition evenly over the folds; further rejection sampling was performed to enforce good statistical qualities of each fold. In particular, the following quantities were optimized by random sampling, in order of importance:

\begin{itemize}
    \item Number of pairs.
    \item Fraction of rounds belonging to high-rapport pairs (class balance at sample level).
    \item Number of rounds (total samples).
    \item Fraction of seconds belonging to high-rapport pairs (class balance at duration level).
    \item Total video length.
\end{itemize}

Each listed quantity was evaluated using the relative square error compared to a uniform distribution over the folds, and weighted according to its importance. This resulted in 5 folds, each containing 7 pairs (3 to 4 pairs per condition). Distributions of game rounds per fold range from 48\% high-rapport to 57\% high-rapport (in total, 21 to 22 rounds per fold).

\subsection*{Model Selection}
\label{sec:ml:model-selection}

Three shallow ML models were chosen, due to their interpretability, and to set a statistically sound baseline for further study: logistic regression, Support Vector Machines (SVM), and decision trees. We performed the subsequent analysis in Python, using scikit-learn\footnote{\url{https://scikit-learn.org/}}~\cite{pedregosa2011scikit}. A set of hyperparameter values was identified for each model; the hyper-parameters and their tested values are shown in Table \ref{tab:hyperparams}.

For each possible combination of video source, feature set, and classification model, we performed nested $k$-folds cross-validation. The inner 4-fold cross-validation loop was used to perform a grid search and obtain optimal hyper-parameters. The outer 5-fold cross-validation loop was used to estimate the model's performance. To be precise, the following procedure was followed:

\begin{enumerate}
    \item For each of the five pre-determined folds:
    \begin{enumerate}
        \item The fold is marked as the test set.
        \item For each remaining fold:
        \begin{enumerate}
            \item The fold is marked as the validation set. The remaining three folds are marked as the train set.
            \item Each combination of hyper-parameters (from the valid combinations shown in Table \ref{tab:hyperparams}) is used to train on the train set, and evaluated on the validation set.
        \end{enumerate}
        \item The best performing hyper-parameters are chosen (based on average accuracy on the validation sets), and evaluated on the test set.
    \end{enumerate}
    \item The expected accuracy of the classification model is estimated as the average test accuracy.
\end{enumerate}

\begin{table}[!ht]
    \caption{Hyperparameters optimized for each model type. All valid combinations were tested in a grid search (e.g., SVM's gamma parameter only applies to the radial basis function kernel).}
    \label{tab:hyperparams}
    \begin{center}
        \begin{tabular}{|lll|}
            \hline
            \textbf{model} & \textbf{hyperparameters} & \textbf{tested values} \\
            \thickhline
            logistic & solver  & \{ lbfgs, saga \} \\
                   & penalty & \{ None, L1, L2, elasticnet \} \\
                   & C       & $10^\lambda$, $\lambda \in \{-2, -1.\hat{6}, \ldots, +2\}$ \\
            \hline
            SVM & C      & $10^\lambda$, $\lambda \in \{-2, -1.\hat{6}, \ldots, +2\}$ \\
                & kernel & \{ linear, rbf \} \\
                & gamma  & \{ scale, auto \} \\
            \hline
            tree & criterion & \{ gini, entropy \} \\
                 & max depth & \{ 2, 3, 4, 5 \} \\
            \hline
        \end{tabular}
    \end{center}
\end{table}

\subsection*{Single-Child Baseline}
\label{sec:ml:single-child}

In the single-child baseline, summary statistics for one child (calculated over the AU data from one game round) are used as features to predict the pair-level label (high-rapport vs. low-rapport). This means we obtain two separate samples from one game round, for a total of 212 samples. Feature sets range from 1 to 4 features per sample. Combining the 25 feature sets described in \emph{\nameref{sec:ml:feature-selection}} and the three models described in \emph{\nameref{sec:ml:model-selection}}, we obtain 75 separate experiments for the single-child baseline.

Compared to the baseline random classifier accuracy of 53.8\%, all 75 feature-model combinations trained on the \textit{left camera} obtained better-than-random accuracy on the train set, and 64/75 obtained better-than-random accuracy on the test set. Respectively for the \textit{right camera}, 74/75 models performed better-than-random on the train set, and 58/75 performed better-than-random on the test set. These high success rates suggest that the AU features generally contain rapport information, and some amount can be extracted with any simple ML strategy.

Table \ref{tab:ml:best-single-child-models-left} lists the top 10 models, ranked by test accuracy on the \textit{left camera}. Two entries tie for best performance: training a decision tree using AU10 and AU12 presence with either the mean-standard deviation combination or the 95\% quantile yields a test accuracy of 68.40\%. Most listed models rely on AU10, which was selected using the data-driven approach. Comparing model performance between the two camera streams (rightmost columns on the table), there seems to be a substantial generalization gap, with few models performing comparatively well on both sources.

\begin{table}[!ht]
\begin{adjustwidth}{-2.25in}{0in} 
    \caption{Top 10 single-child models ranked by \textit{left camera} test accuracy. Listed values include the left camera's test accuracy, and the right camera's test accuracy. q95 indicates 95\% quantile; std indicates standard deviation.}
    \label{tab:ml:best-single-child-models-left}
    \begin{center}
        \begin{tabular}{|llll|cc|}
            \hline
\textbf{model} & \textbf{AUs} & \textbf{statistic} & \textbf{type} & \textbf{left camera} & right camera \\
\thickhline
decision tree & AU10, AU12 & mean, std &  presence & $\mathbf{68.40\% \scriptstyle{\pm 6.16\%}}$ & $55.71\% \scriptstyle{\pm  9.00\%}$ \\
decision tree & AU10, AU12 &       q95 &  presence & $\mathbf{68.40\% \scriptstyle{\pm 6.60\%}}$ & $63.33\% \scriptstyle{\pm 14.15\%}$ \\
          SVM &       AU10 &       q95 &  presence & $66.06\% \scriptstyle{\pm 4.66\%}$ & $62.38\% \scriptstyle{\pm 13.08\%}$ \\
decision tree &       AU10 &       q95 &  presence & $66.06\% \scriptstyle{\pm 4.66\%}$ & $61.90\% \scriptstyle{\pm 13.98\%}$ \\
       linear &       AU10 &       q95 &  presence & $66.06\% \scriptstyle{\pm 4.66\%}$ & $61.90\% \scriptstyle{\pm 13.98\%}$ \\
          SVM & AU10, AU12 &       q95 &  presence & $65.11\% \scriptstyle{\pm 4.98\%}$ & $57.62\% \scriptstyle{\pm 12.19\%}$ \\
       linear & AU10, AU12 &       q95 &  presence & $65.11\% \scriptstyle{\pm 5.52\%}$ & $57.62\% \scriptstyle{\pm 12.19\%}$ \\
          SVM & AU12, AU26 &       q95 & intensity & $64.24\% \scriptstyle{\pm 7.96\%}$ & $57.62\% \scriptstyle{\pm 12.30\%}$ \\
       linear & AU12, AU25 & mean, std &  presence & $64.22\% \scriptstyle{\pm 8.97\%}$ & $52.86\% \scriptstyle{\pm  4.26\%}$ \\
          SVM &       AU12 &       q95 & intensity & $63.79\% \scriptstyle{\pm 8.85\%}$ & $61.90\% \scriptstyle{\pm  5.32\%}$ \\
            \hline
        \end{tabular}
    \end{center}
\end{adjustwidth}
\end{table}

Table \ref{tab:ml:best-single-child-models-right} lists the top 10 models, ranked by test accuracy on the \textit{right camera}. The best performing model is a decision tree trained using the AU12 intensity 95\% quantile, with a train accuracy of 69.77\%, and a test accuracy of 65.71\%. A theory-based combination ranks second; all listed solutions rely on AU12. While overall performance is slightly lower than in the left camera leaderboard, the generalization gap is also smaller between camera sources.

\begin{table}[!ht]
\begin{adjustwidth}{-2.25in}{0in} 
    \caption{Top 10 single-child models ranked by\textit{ right camera} test accuracy. Listed values include the left camera's test accuracy, and the right camera's test accuracy. q95 indicates 95\% quantile; std indicates standard deviation.}
    \label{tab:ml:best-single-child-models-right}
    \begin{center}
        \begin{tabular}{|llll|cc|}
            \hline
\textbf{model} & \textbf{AUs} & \textbf{statistic} & \textbf{type} & left camera & \textbf{right camera} \\
\thickhline
decision tree &       AU12 &       q95 & intensity & $54.76\% \scriptstyle{\pm 5.83\%}$ & $\mathbf{65.71\% \scriptstyle{\pm  3.98\%}}$ \\
       linear & AU06, AU12 &       q95 & intensity & $57.10\% \scriptstyle{\pm 1.76\%}$ & $64.29\% \scriptstyle{\pm  9.07\%}$ \\
       linear &       AU12 &       q95 & intensity & $62.36\% \scriptstyle{\pm 9.61\%}$ & $63.81\% \scriptstyle{\pm  7.97\%}$ \\
decision tree & AU12, AU26 &       q95 &  presence & $63.70\% \scriptstyle{\pm 7.06\%}$ & $63.33\% \scriptstyle{\pm  8.52\%}$ \\
          SVM & AU12, AU26 &       q95 &  presence & $63.70\% \scriptstyle{\pm 7.06\%}$ & $63.33\% \scriptstyle{\pm  8.52\%}$ \\
       linear & AU12, AU26 &       q95 &  presence & $62.27\% \scriptstyle{\pm 4.90\%}$ & $63.33\% \scriptstyle{\pm  8.52\%}$ \\
decision tree & AU10, AU12 &       q95 &  presence & $68.40\% \scriptstyle{\pm 6.60\%}$ & $63.33\% \scriptstyle{\pm 14.15\%}$ \\
decision tree & AU10, AU12 &       q95 & intensity & $52.40\% \scriptstyle{\pm 8.40\%}$ & $62.86\% \scriptstyle{\pm  6.43\%}$ \\
decision tree & AU12, AU25 &       q95 & intensity & $54.70\% \scriptstyle{\pm 6.49\%}$ & $62.38\% \scriptstyle{\pm  5.16\%}$ \\
       linear & AU12, AU26 &       q95 & intensity & $62.79\% \scriptstyle{\pm 6.51\%}$ & $62.38\% \scriptstyle{\pm  7.02\%}$ \\
            \hline
        \end{tabular}
    \end{center}
\end{adjustwidth}
\end{table}

\begin{figure}[!h]
   \centering
   \includegraphics[width=\linewidth]{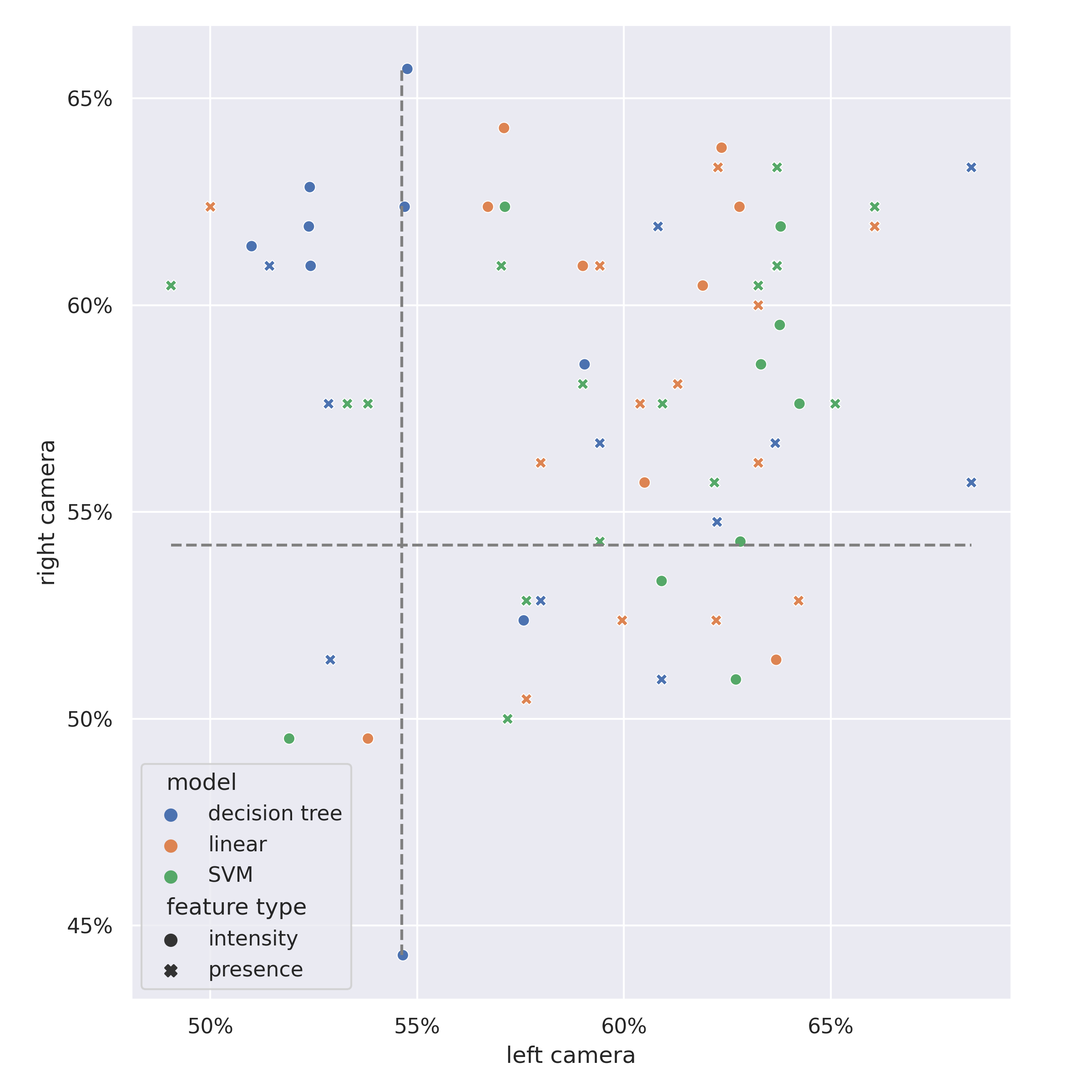}
   \caption{Performance comparison of each single-child model across video sources. Axes indicate the test accuracies in the \textit{left camera} (horizontal axis) and \textit{right camera} (vertical axis). Color indicates model type; shape indicates feature type. Gray dashed lines indicate the random chance baseline.}
   \label{fig:ml:single-child-left-vs-right-test-acc}
\end{figure}

Figure \ref{fig:ml:single-child-left-vs-right-test-acc} shows a comparison of the single-child test accuracies across both camera streams. We can see there is no strong preference for either of the three models, and the best-performing AU estimate type depends on the evaluated source.

\subsection*{Joint Pair Baseline}
\label{sec:ml:pair}

In the joint pair baseline, the AU summary statistics from both children in a game round are concatenated together. Compared to the \emph{\nameref{sec:ml:single-child}}, this means the feature count is doubled to 2-8 features per sample, and the sample count is halved to 106 samples. We use the same methodology, including the same choice of feature sets and ML models, again adding up to 75 feature-model combinations per video source.

All 75 \textit{left camera} feature-model combinations obtained better-than-random train accuracy, and 64/75 obtained better-than-random test accuracy. Respectively for the \textit{right camera}, 73/75 model-feature combinations obtain better-than-random train accuracy, and 54/75 obtain better-than-random test accuracy. These numbers are similar to the ones obtained in single-child experiments.

Table \ref{tab:ml:best-pair-models-left} lists the top 10 joint pair models, ranked by test accuracy on the \textit{left camera}. Again, we get a tie: training either a decision tree or an SVM on the AU10 presence 95\% quantile yields a test accuracy of 70.74\%. While the model choice doesn't seem to matter much, most entries rely on AU10 presence estimates, with the notable exception of a theory-based result. Test accuracies are generally higher than in the single-child baseline, but standard deviations are larger --- suggesting that the sample size reduction causes some instability.

\begin{table}[!ht]
\begin{adjustwidth}{-2.25in}{0in} 
    \caption{Top 10 joint pair models ranked by \textit{left camera} test accuracy. Listed values include the left camera's test accuracy, and the right camera's test accuracy. q95 indicates 95\% quantile; std indicates standard deviation.}
    \label{tab:ml:best-pair-models-left}
    \begin{center}
        \begin{tabular}{|llll|cc|}
            \hline
\textbf{model} & \textbf{AUs} & \textbf{statistic} & \textbf{type} & \textbf{left camera} & right camera \\
\thickhline
decision tree &       AU10 &       q95 &  presence & $\mathbf{70.74\% \scriptstyle{\pm 10.90\%}}$ & $61.90\% \scriptstyle{\pm 17.50\%}$ \\
          SVM &       AU10 &       q95 &  presence & $\mathbf{70.74\% \scriptstyle{\pm 10.90\%}}$ & $61.90\% \scriptstyle{\pm 17.50\%}$ \\
       linear &       AU10 &       q95 &  presence & $68.83\% \scriptstyle{\pm 13.38\%}$ & $61.90\% \scriptstyle{\pm 17.50\%}$ \\
       linear & AU10, AU12 &       q95 &  presence & $65.97\% \scriptstyle{\pm 10.52\%}$ & $64.76\% \scriptstyle{\pm 17.37\%}$ \\
decision tree & AU06, AU12 &       q95 & intensity & $65.06\% \scriptstyle{\pm 11.05\%}$ & $65.71\% \scriptstyle{\pm  9.16\%}$ \\
decision tree &       AU12 & mean, std &  presence & $64.98\% \scriptstyle{\pm  9.01\%}$ & $49.52\% \scriptstyle{\pm  7.97\%}$ \\
          SVM &       AU25 & mean, std &  presence & $64.20\% \scriptstyle{\pm  7.60\%}$ & $49.52\% \scriptstyle{\pm  7.97\%}$ \\
       linear & AU10, AU12 & mean, std &  presence & $64.11\% \scriptstyle{\pm  8.80\%}$ & $56.19\% \scriptstyle{\pm 12.33\%}$ \\
       linear & AU12, AU26 &       q95 & intensity & $64.11\% \scriptstyle{\pm  9.42\%}$ & $58.10\% \scriptstyle{\pm  6.21\%}$ \\
          SVM & AU10, AU12 &       q95 &  presence & $64.07\% \scriptstyle{\pm  8.95\%}$ & $65.71\% \scriptstyle{\pm 16.97\%}$ \\
            \hline
        \end{tabular}
    \end{center}
\end{adjustwidth}
\end{table}

Table \ref{tab:ml:best-pair-models-right} lists the top 10 joint pair models, ranked by test accuracy on the \textit{right camera}. The best-performing model is a decision tree trained using a pure theory-based approach: AU06 and AU12 mean-standard deviation combination. It attains a train accuracy of 81.43\%, and a test accuracy of 70.48\%. Again, we observe higher accuracies but larger standard deviations when compared to single-child models.

\begin{table}[!ht]
\begin{adjustwidth}{-2.25in}{0in} 
    \caption{Top 10 joint pair models ranked by \textit{right camera} test accuracy. Listed values include the left camera's test accuracy, and the right camera's test accuracy. q95 indicates 95\% quantile; std indicates standard deviation.}
    \label{tab:ml:best-pair-models-right}
    \begin{center}
        \begin{tabular}{|llll|rr|}
            \hline
\textbf{model} & \textbf{AUs} & \textbf{statistic} & \textbf{type} & left camera & \textbf{right camera} \\
\thickhline
decision tree & AU06, AU12 & mean, std & intensity & $57.53\% \scriptstyle{\pm 12.17\%}$ & $\mathbf{70.48\% \scriptstyle{\pm 12.78\%}}$ \\
          SVM & AU12, AU26 &       q95 &  presence & $56.67\% \scriptstyle{\pm  7.97\%}$ & $68.57\% \scriptstyle{\pm 10.96\%}$ \\
       linear & AU06, AU12 &       q95 & intensity & $61.26\% \scriptstyle{\pm  6.58\%}$ & $67.62\% \scriptstyle{\pm 15.58\%}$ \\
decision tree &       AU12 &       q95 &  presence & $61.21\% \scriptstyle{\pm  9.54\%}$ & $66.67\% \scriptstyle{\pm 11.17\%}$ \\
       linear & AU12, AU26 &       q95 &  presence & $59.31\% \scriptstyle{\pm  9.72\%}$ & $66.67\% \scriptstyle{\pm 11.17\%}$ \\
          SVM &       AU12 &       q95 &  presence & $56.67\% \scriptstyle{\pm  7.97\%}$ & $66.67\% \scriptstyle{\pm 11.17\%}$ \\
       linear & AU06, AU12 &       q95 &  presence & $54.68\% \scriptstyle{\pm  6.66\%}$ & $65.71\% \scriptstyle{\pm 11.37\%}$ \\
decision tree & AU12, AU26 &       q95 &  presence & $61.21\% \scriptstyle{\pm  9.54\%}$ & $65.71\% \scriptstyle{\pm 11.86\%}$ \\
decision tree & AU12, AU25 &       q95 &  presence & $60.30\% \scriptstyle{\pm  8.31\%}$ & $65.71\% \scriptstyle{\pm 11.86\%}$ \\
          SVM & AU12, AU25 &       q95 &  presence & $55.76\% \scriptstyle{\pm  7.75\%}$ & $65.71\% \scriptstyle{\pm 11.86\%}$ \\
            \hline
        \end{tabular}
    \end{center}
\end{adjustwidth}
\end{table}

\begin{figure}[!h]
   \centering
   \includegraphics[width=\linewidth]{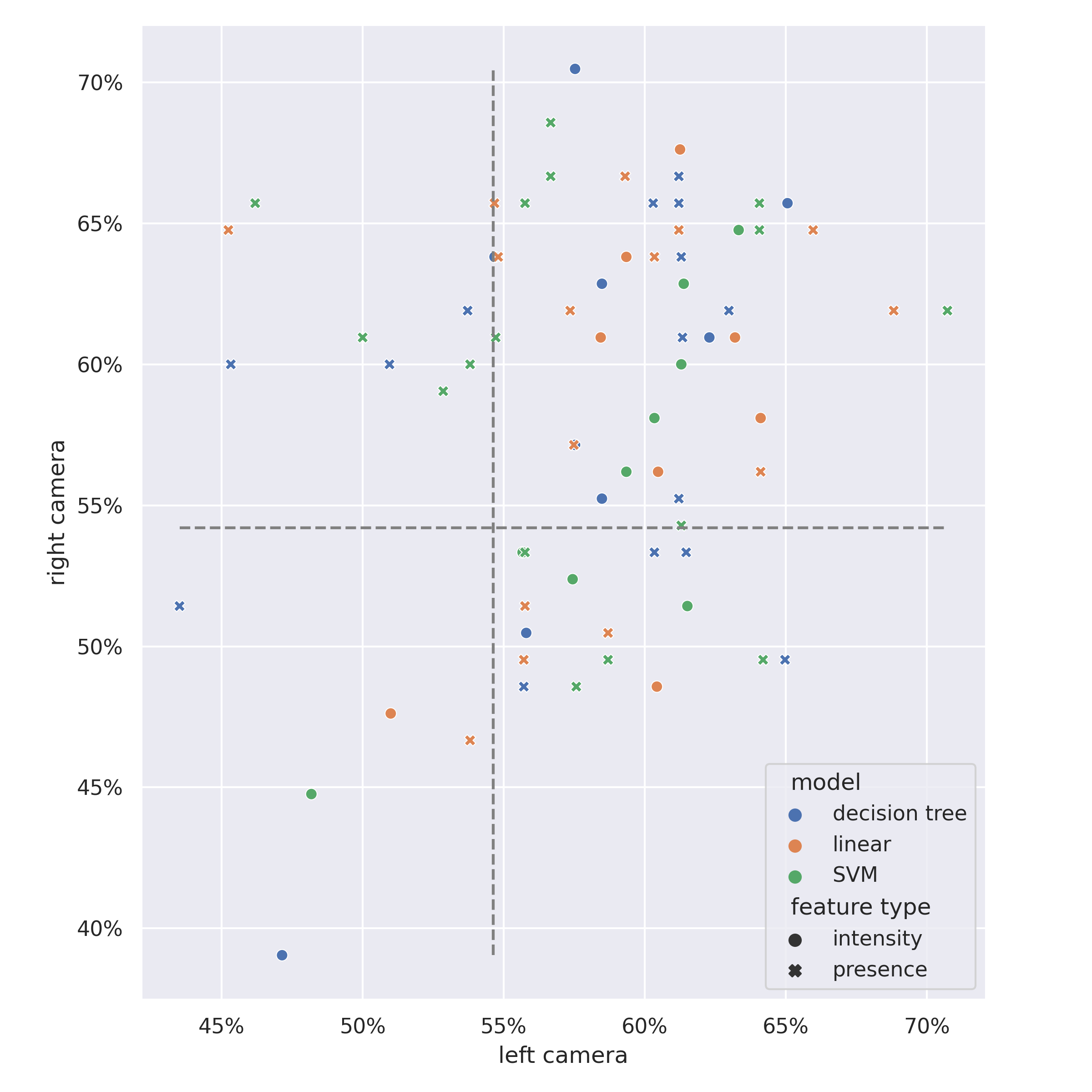}
   \caption{Performance comparison of each joint-pair model across video sources. Axes indicate the test accuracies in the \textit{left camera} (horizontal axis) and \textit{right camera} (vertical axis). Color indicates model type; shape indicates feature type. Gray dashed lines indicate the random chance baseline.}
   \label{fig:ml:joint-pair-left-vs-right-test-acc}
\end{figure}

Figure \ref{fig:ml:joint-pair-left-vs-right-test-acc} shows a comparison of the joint pair test accuracies across both camera streams. We can see a similar overall distribution as in Figure \ref{fig:ml:single-child-left-vs-right-test-acc}, with higher overall accuracies reported.

Overall, using joint data from both children resulted in slightly better accuracies, but noticeably higher standard deviations. In this scenario, taking the interpersonal aspect of rapport into account did not outweigh the benefits from having a bigger population size.

\section*{Discussion and Conclusions}

To better understand the social dynamics and behaviors that emerge in the classroom when children interact with their friends and peers, we collected UpStory: a new dataset of dyadic interactions with different levels of rapport. In order to provide an objective measure of rapport, we leveraged \textit{friendship network} analysis to propose and validate a novel pair-making technique. Our approach allowed us to control for different levels of rapport, resulting in a multimodal dataset on child-child interactions with \textit{high-rapport} and \textit{low-rapport} levels.   

We recorded pairs of children participating in collaborative storytelling play for a total of 35 sessions, adding up to 106 game rounds, with a duration of 3h 40m. Each session is annotated with the pair's condition (high-rapport or low-rapport) and a \textit{social distance heuristic}, allowing practitioners to train ML models based on the children's self-reported friendships. The resulting private dataset contains three video feeds with associated audio, and two additional audio feeds from head-mounted microphones, providing clean recordings for each child's voice. UpStory is the anonymized dataset obtained by extracting frame-by-frame body and face features from two video sources; it is made publicly available at \url{https://zenodo.org/doi/10.5281/zenodo.12635620}. The provided features were extracted using OpenPose and OpenFace, with a custom solution for identity tracking over time. 

We followed two methodologies to validate our pair-making technique. Firstly, analysis of the \textit{social distance heuristic}'s distribution suggested that the pairing strategy worked at the population level, but not at the case-by-case level. This was expected, since we optimized for the sum of distances, but it is a limitation when strong labels are required. A possibility for further ML experiments would be to base the labels on the pair distances instead of the assigned condition, although this can introduce confounding variables (e.g., sociable children being over-represented in the low-distance case). We observed better quality pairings in the Year 2 cohort, both due to a bigger cohort size and due to the participation of several classes. A possible extension to this method is to ensure even participation across several classes; in our study, we were limited by the requirement that the students speak English fluently.

Secondly, our questionnaire analysis showed that children felt significantly closer to their pair in the high-rapport condition, supporting H1. However, our survey data also revealed that children did not necessarily have more positive emotions when playing in the high-rapport condition than in the low-rapport condition, rejecting H2. This result is not surprising, as the novelty effect of the activity could lead to the observed ceiling effects when questionnaires are used to capture children's subjective perceptions~\cite{ligthart2018challenges}. 

Overall, we demonstrated that our proposed pair-making technique is reliable in controlling rapport levels. We believe that our experimental manipulation captures more naturalistic behaviors inherent to the social dynamics of children's dyads providing a more reliable level of rapport than post-hoc annotations encountered in the literature.

The naturalistic social dynamics captured by our dataset hold promise for the training of ML-based automatic prediction of rapport. To build on this, we have provided ML baselines for the prediction of the experimental condition. Both baselines using the data from a single child and baselines using joint pair data are provided. In all cases, the best-performing models attained test accuracies in the 60-70\% range, well above random chance.

We hope the UpStory dataset will contribute to the design of data-driven methodologies to further our understanding of children's relationships with their friends and peers, as well as the development of new technologies that interact with children in an educational context. 

\section*{Acknowledgments}

This work was partly funded by the Centre for Interdisciplinary Mathematics, Uppsala University, and the Swedish Research Council (grant n. 2020-03167).

The authors wish to thank all the colleagues who donated time and effort to help complete this project. In no particular order: Alessio Galatolo, Gustaf Gredebäck, Johan Öfverstedt, and Rebecca Stower.

Attribution for the card images displayed in Figure \ref{fig:board}, in reading order: graveyard by upklyak; mummy by upklyak; grim reaper by upklyak; savannah by macrovector; boy scout by brgfx; hearth by macrovector; dragon hoard by upklyak; elf by pch.vector; shield by vectorpocket.

\nolinenumbers

%
%
%







\bibliography{bibliography}

\end{document}